\documentclass[prd,groupedaddress,showpacs,showkeys,onecolumn,nofootinbib]{revtex4}
\usepackage{graphicx}
\usepackage{dcolumn}
\usepackage{amssymb}
\usepackage{bm}
\usepackage{enumerate}
\begin{document}
%----------------------------------------------------------------
\title{Threshold effects and Planck scale Lorentz violation:\\
combined constraints from high energy astrophysics}
%-------------------------------------------------------------------------
\author{T. Jacobson}
\email{jacobson@physics.umd.edu}
%-------------------------------------------------------------------------
\author{S. Liberati}
\email{liberati@physics.umd.edu}
%-------------------------------------------------------------------------
\author{D. Mattingly}
\email{davemm@physics.umd.edu}
%-------------------------------------------------------------------------
\affiliation{Physics Department, University of Maryland, College
Park, MD 20742-4111, USA}
%-------------------------------------------------------------------------
\date{\today}
%-------------------------------------------------------------------------
\bigskip
%-------------------------------------------------------------------------
\begin{abstract}
%-------------------------------------------------------------------------
\bigskip

Recent work has shown that dispersion relations with Planck scale
Lorentz violation can produce observable effects at energies many
orders of magnitude below the Planck energy $M$. This opens a window
on physics that may reveal quantum gravity phenomena. It has already
constrained the possibility of Planck scale Lorentz violation, which
is suggested by some approaches to quantum gravity.  In this work we
carry out a systematic analysis of reaction thresholds, allowing
unequal deformation parameters for different particle dispersion
relations. The thresholds are found to have some unusual properties
compared with standard ones, such as asymmetric momenta for pair
creation and upper thresholds.  The results are used together with
high energy observational data to determine combined constraints. We
focus on the case of photons and electrons, using vacuum \v{C}erenkov,
photon decay, and photon annihilation processes to determine order
unity constraints on the parameters controlling $O(E/M)$ Lorentz
violation. Interesting constraints for protons (with photons or pions)
are obtained even at $O((E/M)^2)$, using the absence of vacuum
\v{C}erenkov and the observed GZK cutoff for ultra high energy cosmic
rays.  A strong \v{C}erenkov limit using atmospheric PeV neutrinos is
possible for $O(E/M)$ deformations provided the rate is high
enough. If detected, ultra high energy cosmological neutrinos might
yield limits at or even beyond $O((E/M)^2)$.

%-------------------------------------------------------------------------
\end{abstract}
%-------------------------------------------------------------------------
%-------------------------------------------------------------------------
%-------------------------------------------------------------------------
\maketitle
%-------------------------------------------------------------------------
% Local defines
%----------------------------------------------------------------
\def\wt{\widetilde}
\def\gsim{\; \raisebox{-.8ex}{$\stackrel{\textstyle >}{\sim}$}\;}
\def\lsim{\; \raisebox{-.8ex}{$\stackrel{\textstyle <}{\sim}$}\;}
\def\half{{1\over2}}
\def\a{\alpha}
\def\b{\beta}
\def\g{\gamma}
\def\d{\delta}
\def\e{\epsilon}
\def\o{\omega}
\def\m{\mu}
\def\t{\tau}
\def\L{{\mathcal L}}
\def\p{{\mathbf{p}}}
\def\q{{\mathbf{q}}}
\def\k{{\mathbf{k}}}
\def\fp{{p_{\rm 4}}}
\def\fq{{q_{\rm 4}}}
\def\fk{{k_{\rm 4}}}
\def\etal{{\emph{et al}}}
\def\det{{\mathrm{det}}}
\def\tr{{\mathrm{tr}}}
\def\ie{{\emph{i.e.}}}
\def\aka{{\emph{aka}}}
%----------------------------------------------------------------
\def\HRULE{{\bigskip\hrule\bigskip}}
%----------------------------------------------------------------
%----------------------------------------------------------------
%----------------------------------------------------------------
\section{Introduction}
%----------------------------------------------------------------
The principle of relativity of motion goes all the way back to
Galileo~\cite{dialog}, who noted that observers below decks in a large
ship gliding across a calm sea have no way of determining whether they
are in motion or at rest. Einstein's special relativity, which is
founded on this principle, has been spectacularly successful in
accounting for phenomena involving boost factors as high as
$10^{11}$. Moreover, the Lorentz group has a beautiful mathematical
structure, and this symmetry powerfully constrains theories in a way
that has been very useful in discovering new laws of physics. It is
natural to assume under these circumstances that Lorentz invariance is
a symmetry of nature up to arbitrary boosts. Nevertheless, there are
several good reasons to question exact Lorentz symmetry. ~From a
logical point of view, the most compelling reason is that while
$10^{11}$ is a large number, it is nowhere near infinity. There is,
and will always be, an infinite volume of the Lorentz group that is
experimentally untested since, unlike the rotation group, the Lorentz
group is non-compact. Why should we assume that {\it exact} Lorentz
invariance holds when this hypothesis cannot even in principle be
tested?

While the non-compactness reason for questioning Lorentz symmetry is
perhaps logically compelling, it is by itself not very
encouraging. However, there are also several reasons to suspect that
there will be a failure of Lorentz symmetry at some energy or
boosts. One reason is the ultraviolet divergences of quantum field
theory, which are a direct consequence of the assumption that the
spectrum of field degrees of freedom is boost invariant. Another
reason comes from quantum gravity. Profound difficulties associated
with the ``problem of time'' in quantum gravity~\cite{Isham,Kuchar}
have suggested that an underlying preferred time may be necessary to
make sense of this physics.  Also tentative results in string
theory~\cite{KS89}, quantum geometry~\cite{loopqg}, and
non-commutative geometry~\cite{Hayakawa,Carroll:2001ws,
Amelino-Camelia:2001cm} approaches to quantum gravity have suggested
that Lorentz symmetry may be broken in the ground state.

Finally, there have been recent hints from high energy astroparticle
physics that we may already be seeing the effects of Lorentz violation
(although as discussed below the most recent analyses make this seem
unlikely.)  One comes from the photo-production of electron-positron
pairs when cosmic gamma rays collide with photons of the infrared
background. Below 10 TeV the (indirectly) observed absorption of such
gamma rays by this process offers support for boost invariance up to
the boost that relates the cosmic rest frame to the center of mass
frame of the colliding photons. (For a 10 TeV gamma ray colliding head
on with a 25 meV infrared photon this yields a boost of $10^7$.)
However, according to some (but not all) models of the infrared
background, there appears to be less absorption than expected for
gamma rays above 10 TeV coming from the blazar Mkn 501 (located at
about $157$ Mpc from us). If true this could be explained by an upward
threshold shift due to a Planck scale suppressed Lorentz violating
term in the dispersion relation for the gamma
rays~\cite{Protheroe:2000hp}.

The other hint comes from the cosmic ray events beyond the GZK
cutoff~\cite{G,ZK} on high energy protons. Ultra high energy protons
undergo inelastic collisions with CMBR photons leading to the
production of pions (the boost to the center of mass frame yields the
figure of $10^{11}$ mentioned above). As a result, protons above $\sim
5\times10^{19}$ eV are not able to reach us from distances above a few
Mpc~\cite{Stecker68}. In spite of this prediction, cosmic rays with
energy beyond $10^{20}$ eV have apparently been observed by the AGASA
experiment~\cite{GZKdata} (see also~\cite{NW00} for a review on this
issue). The nature and origin of these ultra high energy cosmic rays
is unknown and several explanations have been proposed
(see~\cite{Sigl, Stecker:2002fh} for an extensive review). One
proposal is that Lorentz violating terms in the dispersion relation
for the proton produce an upward shift of the threshold for pion
production, allowing these high energy protons to reach
us~\cite{Mestres,CG,Bertolami,ACP}. Interestingly it was argued that a
universal Lorentz violating deformation of the particle dispersion
relations would be capable of explaining both the TeV gamma ray
absorption anomaly and the trans-GZK events~\cite{ACP}.

The evidence for the TeV gamma ray and GZK anomalies is not convincing
at this stage, however. Indeed it has been argued in~\cite{stecker01,
Stecker:2002fh} for the former and in~\cite{Bahcall:2002wi,FLYeye02}
for the latter that the data are consistent with Lorentz
invariance. For us therefore the most important point is just that it
is possible at all that Planck scale violations of Lorentz symmetry
could be observed or constrained by current and upcoming
observations. The focus of the present paper is almost entirely on the
{\it constraints} that can be imposed. Our work extends prior
results~\cite{Mestres,CG,Bertolami,Acea,ACP,Kluzniak,Kifune:1999ex,Aloisio:2000cm}
in several ways: (i) combining constraints to limit parameter space of
{\it a priori} independent parameters, (ii) discovery and
characterization of the asymmetric threshold effect, (iii)
characterization of upper threshold effects, (iv) extending analysis
for threshold effects to higher order nonlinearities.  A brief report
on some of our results has already been given
in~\cite{Jacobson:2001tu}. Some of these results have been confirmed
in~\cite{Major}.

In the next section we discuss our theoretical framework and list the
reactions we are going to consider. In Section~\ref{sec:qed} we study
the kinematics of some photon--electron processes in order to
determine how Lorentz violating dispersion affects thresholds. The
details of the photon annihilation threshold analysis are worked out
in the Appendix. These results are then used to deduce observational
constraints on the electron and photon deformation parameters.  Taken
jointly these constraints severely restrict the parameter
plane. Section~\ref{sec:other} is devoted to the discussion of other
possible interactions including hadrons or neutrinos, and in
section~\ref{sec:univ} we discuss the special case of common Lorentz
violating parameters for all the particles.  Finally we present some
conclusions and perspectives in section~\ref{sec:disc}.

Throughout this paper we adopt the following notational
conventions: $\fp$ denotes a four-momentum $\fp=(\omega,\p)$, and
$p$ is the magnitude of the three-vector $\p$. The metric
signature is $(+,-,-,-)$. We use the energy scale $M=10^{19}$ GeV
to form dimensionless Lorentz-violating parameters, since it is
close to the Planck energy $M_{\rm P}=(\hbar c^5/G)^{1/2}\simeq 1.22
\cdot10^{19}$ which we are presuming sets the scale for violation
of Lorentz invariance induced by quantum gravity. We often employ
units in which $M=1$.

%---------------------------------------------------------
\section{Theoretical framework and processes considered}
\label{sec:fram}
%---------------------------------------------------------

Various approaches to quantum gravity have suggested that violations
of local Poincar\'e symmetry might occur, but no reliable prediction
is currently availble.  These suggestions range from the breaking of
just the boost symmetry to breaking of the full local Poincar\'e
group.  In this paper we study the former case since it is the minimal
one for which consequences of boost symmetry violation can be
explored. Thus we shall assume that rotation and spacetime translation
symmetries are preserved, so that in particular energy and momentum
are conserved.\footnote{For an example where both rotation and boost
symmetry are broken see e.g.~\cite{Kc}. For an exploration of the case
in which the full Poincar\'e symmetry is violated see
e.g.~\cite{Ellis-etal,Ng:1993jb}.}

Dispersion relations determine how particles propagate and, via
energy-momentum conservation, how their interactions are kinematically
constrained. Hence Lorentz violating dispersion relations provide a
relatively theory-independent window into the possibility of Lorentz
violating physics. In this work we explore the observational
consequences of such deformed dispersion relations in flat spacetime,
i.e neglecting gravitational effects.  The consequences of such
dispersion relations have also been extensively investigated in the
context of the Hawking effect (see e.g. \cite{Jacobson:1999zk} and
references therein) and the primordial spectrum of density
fluctuations in cosmology (see
e.g.~\cite{Niemeyer:2002ze,Niemeyer:2002kh,Brandenberger:2002sr} and
references therein).
  
In this section we discuss our framework for parametrizing such
Lorentz violating physics, as well as the processes through which one
might hope to place constraints or to observe Lorentz violation.

%---------------------------------%
\subsection{Theoretical framework}
%---------------------------------%

A dispersion relation that is not boost invariant can hold in only one
frame. We assume this frame coincides with that of the cosmic
microwave background.  As mentioned above, we further assume that
rotation invariance is preserved in this preferred frame.  Thus the
dispersion relation takes the form $E=E(p)$, where $p$ is the
magnitude of $\p$. In the Lorenz invariant case we have $E^2=m^2 +
p^2$.  Effective field theory suggests that it should suffice to
consider generalizations of this form involving only integer powers of
momentum,
\begin{equation}
E^2
= m^2 + p^2 + \sum_{n=1}^{\infty} a_n p^n.
\label{eq:drel}
\end{equation}

We presume that any Lorentz violation is associated with
quantum gravity and suppressed by at least one inverse
power of the Planck scale $M$.  For $n\ge3$ it is
therefore natural to factor out the appropriate power of
$M$ and write $a_n=\eta_n/M^{n-2}$ where $\eta_n$ is a
dimensionless constant that might be expected to be of
order unity if indeed quantum gravity does violate
Lorentz symmetry. For $n<3$ there must in addition be
another mass scale, $\mu$, which might be a particle
physics mass scale, in terms of which the coefficents
$a_{1,2}$ can be written as $a_1=\a_1\, \mu^2/M$ and
$a_2=\a_2\, \mu/M$, where again $\a_{1,2}$ might be
expected to be of order unity.\footnote{ Renormalization
  group arguments might suggest that lower powers of
  momentum in Eq.~(\ref{eq:drel}) will be suppressed by
  lower powers of $M$. However this need not be the case
  if a symmetry or other mechanism protects the lower
  dimension operators from Lorentz violation.  See
  e.g.~\cite{Burgess:2002tb} for an example of this in a
  brane-word scenario where there is Lorentz invariance
  on the brane but not off the brane.} In a situation
such as this, the important terms at low energies
$p\ll\mu$ would be from $p^1$ and $p^2$.  At high
energies $p\gg\mu$, the $p^3$ term if present would
dominate. If this term is absent then the $p^4$ term
would dominate when $p^2\gg\mu M$.

A large amount of both theoretical and experimental work has been done
on the case $n\le2$. The most general framework is the ``standard
model extension''~\cite{Kc}, which includes not just rotation
invariant effects but all possible renormalizable Lorentz and CPT
violating terms that can be added to the standard model Lagrangian
preserving the field content and gauge symmetries. Low energy
observations~\cite{Kc,Kostelecky:Meeting,Kostelecky:2001mb} have
placed stringent limits on the magnitude of such Lorentz and CPT
violating terms. For example in~\cite{Kostelecky:2001mb} a very strong
constraint of order $10^{-32}$ from spectropolarimetry is provided for
the electromagnetic birefringence of the vacuum in the standard model
extension.  High energy astroparticle phenomena~\cite{CG,SG01} have
also been used, however in the case of such phenomena the above
discussion suggests that unless the $p^3$ term is absent it would be
expected to dominate over the $p^2$ and $p^1$
corrections.

In this paper we focus on the constraints that can be obtained from
high energy phenomena. In the absence of peculiar tuning of the
coefficients of the terms with different powers $p^n$, it is natural
to suppose that the lowest nonzero term with $n\ge3$ will dominate at
these energies.  Hence, for simplicity, we shall include only one
Lorentz violating power of momentum. Our study thus amounts to
studying the observational consequences of dispersion relations of the
form
\begin{equation}
 E^2 = p^2+m^{2}_{a} +
%\sum_{n=1}^{\infty}
 \eta_{a} {p^n}/{M^{n-2}}.
 \label{eq:modr}
\end{equation}
The subscript $a$ denotes different particles, and {\it a priori} all
  the dimensionless coefficients $\eta_{a}$ could be different. (For
  notational uniformity we use here $\eta_{1,2}$ rather than the
  coefficients $\a_{1,2}$ defined above.) We assume that, in addition
  to being conserved, energy and momentum add for composite systems in
  the usual way.\footnote{
%-------------------------
Note however that there have been recent proposals in which the
composition law for energy and momentum is also
modified~\cite{AC-DSR,Kow-Glik-DSR,Mag-Smo-DSR,JV-DSR}.
%-------------------------
} It might seem that the effects of such deformations of the
dispersion relation could be important only near the Planck
energy. However, there are at least two types of phenomena for which
this is not the case.

First, for particles that propagate over cosmological distances, small
differences in propagation speed can build up to detectable
time-of-flight differences. Second, thresholds for particle reactions
can be shifted, and thresholds can appear for normally forbidden
processes. These threshold effects can occur at energies many orders
of magnitude below the Planck scale. To see why, note that thresholds
are determined by particle masses, hence if the $p^n$ term is
comparable to the $m^2$ term in (\ref{eq:modr}) one can expect a
significant threshold shift. This occurs at the momentum
\begin{equation}
   p_{\rm dev}\sim\left({m^2 M^{n-2}/\eta}\right)^{1/n},
 \label{eq:equal}
\end{equation}
which gives a rough idea of the energies at which we expect to see
deviations from standard physics. The typical scales for some
different particles if $\eta\sim 1$ are summarized in
Table~\ref{tab:en}.
%------------------------------------------------------------------------
\begin{table}[ht]
 \caption{Typical energies at which one can expect deviations from
  standard kinematics for different particles if $\eta\sim1$ and
  $n=3,4$.  The mass of the neutrino is taken to be $\sim 1$ eV, this
  being the current upper bound on the mass of the lightest
  neutrino.\label{tab:en}} \vspace{0.2cm} \begin{center} \footnotesize
  \begin{tabular}{c||c|c|c} \hline {~~n} &
  \raisebox{0pt}[13pt][7pt]{$p_{\rm dev}$ for $\nu_{\rm e}$} &
  \raisebox{0pt}[13pt][7pt]{$p_{\rm dev}$ for $e^{-}$} &
  \raisebox{0pt}[13pt][7pt]{$p_{\rm dev}$ for $p^{+}$} \\ \hline\hline
  {~~3} & \raisebox{0pt}[13pt][7pt]{$\sim 1$ GeV}&
  \raisebox{0pt}[13pt][7pt]{$\sim 10$ TeV}&
%\raisebox{0pt}[13pt][7pt]{$\sim 10^2$ TeV}&
\raisebox{0pt}[13pt][7pt]{$\sim 1$ PeV}\\
\hline {~~4} & \raisebox{0pt}[13pt][7pt]{$\sim 100$ TeV}&
\raisebox{0pt}[13pt][7pt]{$\sim 100$ PeV}&
%\raisebox{0pt}[13pt][7pt]{$\sim 1$ EeV}&
\raisebox{0pt}[13pt][7pt]{$\sim 3$ EeV}\\
\hline
\end{tabular}
\end{center}
\end{table}
%----------------------------------------------------------------

%------------------------------------------------------------
\subsection{Viability of theoretical framework}
%------------------------------------------------------------

Before considering the observational constraints, a few comments are
in order regarding the viability of the theoretical framework we are
adopting.
\begin{itemize}
\item{\bf Restriction to $p\ll M$ and monotonicity of $E(p)$}\\ We
view the dispersion relation just as the initial terms in a derivative
expansion, so we are assuming nothing about the actual Planck scale
physics. In particular, when $n>2$ and $\eta$ is negative, the right
hand side of the dispersion relation (\ref{eq:modr}) becomes negative
for large enough momenta $\sim|\eta|^{-1/(n-2)} M$. However, we never
use the dispersion relation in this regime where the energy would be
imaginary.  Moreover, it will be important for our threshold analysis
that we restrict attention even further to the regime in which the
dispersion relation is strictly monotonic. As long as $|\eta|$ is not
much larger than unity this will be the case provided the momentum is
below the Planck scale. In fact, we consider only momenta many orders
of magnitude below the Planck scale.
\item{\bf Causality and stability}\\ For positive $\eta$ the
propagation is superluminal at high energies. One might worry that
this would lead to causal paradoxes, however this is not the case,
since the propagation is always forward in time relative to the
preferred frame in which the dispersion relation is specified. For
negative $\eta$ the 4-momentum is spacelike at high energy, hence in a
boosted frame the energy can be less than zero. One might think this
implies that the case with $\eta<0$ is not energetically stable and
hence unviable. This is not so, however, since all energies remain
positive relative to the preferred frame, which is enough to guarantee
stability.\footnote{
%---------------------------------
For an alternative point of view, see~\cite{Kostelecky:rh}.
}
%-----------------------------------------------
%
\item{\bf Dispersion relations for macroscopic systems}\\ The
deformed dispersion relations are introduced for elementary
particles only; those for macroscopic objects are then inferred by
addition. For example, if $N$ particles with momentum $\p$ and
mass $m$ are combined, the total energy, momentum and mass are
$E_{\rm tot}=NE(p)$, ${\bf P}_{\rm tot}=N\p$, and $M_{\rm tot}=Nm$, so
that $E_{\rm tot}^{2}= M_{\rm tot}^{2}+ P_{\rm tot}^{2} +
N^{2-n}\eta P_{\rm tot}^n$ (in units with $M=1$).  The ratio of
the Lorentz violating term to the $P^2$ term is the same as it is
for the individual particles, $\eta p^{n-2}$, hence there is no
observational conflict with standard dispersion relations for
macroscopic objects.
\item{\bf Effective field theory and compatibility with general
relativity}\\ There is no difficulty exporting deformed dispersion
relations to curved spacetime, provided they can be produced by an
effective Lagrangian for a field. In this case, the preferred frame in
which the dispersion relation holds is specified by a unit timelike
vector field, which must be promoted to a dynamical field of the
theory if general covariance is to be
preserved~\cite{Kostelecky:1989jw,Carroll:vb,Gasperini,JM}. In the
cases that $n$ is even, there are obvious Lagrangians that produce the
dispersion relation. For example one can add terms involving extra
powers of the spatial Laplacian, such as $({}^{(3)}{\nabla}^2\phi)^2$
for a scalar field. For odd $n$ there seems to be no local action that
will work for real scalar fields, although for a complex scalar the
term $i\bar{\phi}\partial_t{}^{(3)}{\nabla}^{2}\phi+ {\rm h.c.}$
induces cubic and higher order terms. To induce cubic terms for
spinors one can write for example
$\bar{\psi}{}^{(3)}{\nabla}^{2}\psi$, and for the electromagnetic
field one can write ${\bf B}\cdot {\bf \nabla}\times{\bf E}$ (which
violates parity). This last case yields a sort of Lorentz violation
that emerges from quantum geometry calculations~\cite{loopqg}.  The
Lorentz violating terms in the effective Lagrangians just discussed
have mass dimension greater than four so are not renormalizable.  This
is not a fundamental problem, since we only regard the Lagrangian as
an effective one below some large energy scale, however it raises the
question of naturalness.  For now we take the point of view that there
may be an explanation for the low energy Lorentz symmetry that is not
yet understood.
\end{itemize}

%----------------------------------------------------
\subsection{Processes considered}
%----------------------------------------------------

In order to determine the strongest joint constraints on the a
priori independent coefficients $\eta_{a}$ in (\ref{eq:modr}) one
must identify several processes involving the same types of
particles. We focus most of our attention on the case of photons
and electrons, since the electron mass is light and these
particles interact readily. In this way we are able to obtain
rather strong constraints on the allowed parameter space. We also
consider several other processes some of which presently allow or
will soon allow further interesting constraints to be placed. Here
we summarize all the processes to be considered in the paper and a
few more.
\begin{enumerate}[A {)}]
\item \label{list:g-e} {\bf Photon--electron processes}
\begin{enumerate}
\item \label{list:QEDv} {\em QED vertex interactions}: The basic
QED vertex involves one photon and two electron lines. With all
particles on-shell this vertex is forbidden (for any in-state) by
energy momentum conservation in the usual Lorentz invariant
theory, but it can be allowed by Lorentz violating dispersion. In
particular we consider the following processes:
\begin{enumerate}
\item \label{list:Cerenkov} $e^{-} \rightarrow e^{-}\, \gamma$:
This {\em vacuum \v{C}erenkov effect} is extremely efficient,
leading to an energy loss rate that goes like $E^2$ well above
threshold. Thus any electron known to propagate must lie below the
threshold. We shall also discuss the vacuum \v{C}erenkov effect
for other charged particles and even for neutral particles.
\item\label{list:gd} $\gamma \rightarrow e^{+} \, e^{-}$: The {\em
photon decay} rate goes like $E$ above threshold, so any gamma ray
which propagates over macroscopic distances must have energy below
the threshold.
\item\label{list:pa} $e^{+}\, e^{-} \rightarrow \gamma$ {\em Pair
annihilation} to a single photon can also occur. For cosmological
observations this would be hardly distinguishable from the similar
two-photon pair annihilation and as such it is not presently
helpful in providing observational constraints.
\end{enumerate}
\item \label{list:gg} $\gamma \, \gamma \rightarrow e^{+} \, e^{-}$:
{\em Photon annihilation} occurs in ordinary QED above a certain
threshold, however Lorentz violating dispersion can modify this
threshold in observationally interesting ways and can introduce an
upper threshold. (The related reactions of pair annihilation (into two
photons) and Compton scattering are also modified, however these
effects offer no clear signal that can provide useful constraints.)
\item \label{list:gammasplitting} $\gamma \rightarrow N\gamma$:
{\em Photon splitting} is allowed by energy momentum conservation
in Lorentz invariant QED if all $N+1$ photons have parallel
momenta, but the process does not occur both because the matrix
element vanishes and the phase space volume vanishes. With
modified dispersion the photon four-momenta are no longer null
(and there may be additional Lorentz violating operators that
mediate the process) hence this reaction can occur with a finite
rate. However, we shall see that the rate is too small to be
observable.
\item \label{list:tof} {\em Time of flight constraints}: Non-linearity
in the modified dispersion relation leads to different times of
arrival for photons of different wavelength emitted from the same
event. Such differences can provide an upper bound on the parameter
governing the amount of Lorentz violation for photons, independently of
the parameters for other particles.
\item {\em Vacuum birefringence constraints}. Violations of Lorentz
invariance involving also parity violation can lead to unequal speeds
of propagation for different photon polarizations.  The absence of
such birefringence effects for light (IR-UV) from cosmological sources
has been used to provide constraints of order $10^{-32}$ and $10^{-5}$
for the quadratic~\cite{Kostelecky:2001mb} and cubic
deformations~\cite{Gleiser:2001rm} respectively.
\end{enumerate}
\item\label{list:others} {\bf Other processes}:
\begin{enumerate}
\item {\em Alternative vacuum \v{C}erenkov effects}
\begin{itemize}
\item \label{list:protons} $p^+ \rightarrow p^+ \, \gamma$ or $n
\rightarrow n \, \gamma$: Note that to properly analyze this reaction
the structure of the proton or neutron must be taken into
account. \item \label{list:Neutrino} $\nu \rightarrow \nu \, \gamma$:
Although neutrinos are neutral, they still have a charge structure in
the standard model so can in principle produce vacuum \v{C}erenkov
radiation via the charge radius coupling. Massive neutrinos could also
radiate via the magnetic moment coupling~\cite{Rabi}. The related
process of photon decay to neutrinos, $\gamma\rightarrow
\bar{\nu}\nu$, may also provide an interesting constraint.
\item \label{list:gravitons} Gravitational \v{C}erenkov radiation
will occur if matter moves faster than the phase velocity of
gravitons in vacuum~\cite{Bassett:2000wj}. This effect has been
used, in the special case $n=2$, to place limits on the difference
between the maximum speeds of propagation for gravitons and
photons~\cite{Moore:2001bv}.
\end{itemize}
\item $p^{+}\,\gamma_{\rm CMB}\to p^{+}\,\pi_{0}$: {\em GZK interaction}.
Lorentz violations can change the allowed range of energies for this
reaction.  The confirmation of the standard GZK cutoff can therefore
provide interesting constraints even in the case $n=4$ due to the
tremendously high energy of the most energetic cosmic rays.  Moreover
the highest energy events recorded by AGASA may conceivably be
explained via an upper threshold.
\item {\em Neutron stability--Proton instability}. If the
dispersion relations for the neutron and proton are independently
modified, it is possible to make neutrons stable at high energies.
The highest energy AGASA events could be understood in this manner
if the trans-GZK particles were actually neutrons, hence
suppressing their interaction with the cosmic background
radiation.
\item {\em Neutrino oscillations}.  Non flavor-diagonal Lorentz
violations can produce neutrino oscillations, even for massless
neutrinos~\cite{CGne}.  For quadratic deviations in the dispersion
relation ($n=2$) the constraints from current observation have
been considered in~\cite{CG,CGne,Pak} leading to a constraint on
the difference of speed between electron and muon neutrinos of
about $10^{-22}$.  Constraints for higher order Lorentz violations
have been discussed in~\cite{Brustein}.
\item {\em Anisotropy effects}. The motion of the laboratory with
respect to the preferred frame can produce anisotropic effects.
Limits for the case $n=2$ are discussed
in~\cite{Kostelecky:Meeting}.  Such an effect has recently been
used~\cite{Sudarsky} to show that the Lorentz violation suggested
by quantum geometry calculations is in conflict with current
observations in Hughes--Drever type experiments.
\end{enumerate}
\end{enumerate}

%----------------------------------------------------
\subsection{Observations}
%----------------------------------------------------

To obtain constraints from these reactions we shall consider the
following observations:
\begin{enumerate}
\item Electrons of energy up to $\sim 100$ TeV are inferred via
X-ray synchrotron radiation coming from supernova
remnants~\cite{Koyama,Naito}.
\item Gamma rays up to $\sim 50$ TeV arrive on earth from the Crab
nebula~\cite{Tanimori}.
\item Cosmic gamma rays are absorbed in a manner consistent with
photon annihilation off the IR background with the standard
threshold~\cite{stecker01,Stecker:2002fh}. This inference depends on
incomplete knowledge of the IR background and on assumed properties of
the source spectrum however, so the consistency provides only an
imprecise constraint at present.
\item Different photons emitted by the same gamma ray burst all arrive
at earth within a narrow time interval.  
\item The GZK cut off on UHE cosmic ray protons at $\sim 5\cdot
10^{19}$ eV has been observed~\cite{Bahcall:2002wi, FLYeye02}
(although events at higher energy may have been
detected~\cite{GZKdata}).
\end{enumerate}

Using the electron--photon processes we find strong constraints on the
allowable range for the photon and electron parameters for cubic order
($n=3$) Lorentz violation, while the quartic case ($n=4$) is only
weakly constrained.  Using UHE cosmic ray protons we obtain strong
constraints even for $n=4$.

%----------------------------------------------------------------
\section{Photon--electron processes}
\label{sec:qed}
%----------------------------------------------------------------

In this section we determine the thresholds for some elementary
processes involving just photons and electrons. The fact that
these always involve the same particles will allow us to combine
the constraints provided by the available observations and to
severely restrict the space of the Lorentz violating parameters.
~From here on we adopt units with $M=1$, but occasionally display
the $M$ dependence explicitly.

%-------------------------------------------------------------------------
\subsection{Kinematics of the basic QED vertex}
\label{kinematics}
%-------------------------------------------------------------------------
The processes $e^-\rightarrow e^-\gamma$ and $\gamma\rightarrow
e^+e^-$ correspond to the basic QED vertex, but are normally
forbidden by energy-momentum conservation together with the
standard dispersion relations. When the latter are modified, these
processes can be allowed.

For photons and electrons the assumed dispersion relations are:
\begin{eqnarray}
 \omega^2(k)&=& k^2+\xi k^n,
 \label{eq:pdr}\\
 E^2(p)&=& m^2+ p^2+\eta p^n,
 \label{eq:mdr}
\end{eqnarray}
where we have introduced the notation $\xi=\eta_{\gamma}$ and
$\eta=\eta_{e}$. Let us denote the photon 4-momentum by
$k_{4}=(\omega_{k},\k)$, and the electron and positron 4-momenta
by $p_{4}=(E_{p},\p)$ and $q_{4}=(E_{q},\q)$. For the two
reactions energy-momentum conservation then implies $p_{4}=k_{4}+
q_{4}$ and $k_{4}=p_{4}+q_{4}$ respectively. In both cases, we
have
\begin{equation}
(p_{4}-k_{4})^2=q_{4}^2, \label{eq:squares}
\end{equation}
where the superscript ``2" indicates the Minkowski squared norm.
Using the Lorentz dispersion relations (\ref{eq:pdr}) and
(\ref{eq:mdr}) this becomes
\begin{equation}
\xi k^n+\eta p^n-\eta q^n =
2\left(E_{p}\omega_{k}-pk\cos\theta\right), \label{Econs}
\end{equation}
where $\theta$ is the angle between $\p$ and $\k$. In the standard
case the coefficients $\xi$ and $\eta$ are zero and the r.h.s.\ of
Eq.~(\ref{Econs}) is always positive, hence there is no solution.
It is clear that non-zero $\xi$ and $\eta$ can change this
conclusion and allow these processes.

We define a {\em lower threshold} as the minimum energy required
for the incoming particle for the reaction to occur. (If the
initial state is a two particle state, then a threshold is defined
relative to a fixed energy for the ``target'' particle.
Conversely, an {\em upper threshold} is defined as the maximum
energy (if any) allowed for the incoming particle for the reaction
to occur. Our analysis is based on properties of thresholds
summarized in the following threshold theorem:
\begin{quote}
    {\bf Threshold theorem}: {\em If $E_{p}$ is a strictly
monotonically increasing function of $p$ for $p>0$ for all
particles, then all thresholds for processes with two particle
final states occur when the final momenta are parallel. For
processes with two particle initial states the initial momenta at
threshold are anti-parallel.}
\end{quote}
A detailed proof can be found in~\cite{JLMth}. According to the
theorem, $\theta=0$ at a threshold. This point has been assumed in
previous work but was not shown explicitly and in fact is not true
if $E_{p}$ is not monotonic.

Fixing $\theta$ to be zero, all three spatial momenta are
parallel, hence momentum conservation implies
$q=|\q|=|\pm(\p-\k)|=|p-k|$. In this case the relation
(\ref{Econs}) becomes
\begin{equation}
\xi k^n+\eta p^n-\eta |p-k|^n =
2pk\left(\frac{E_{p}}{p}\frac{\omega}{k}-1\right).
\label{eq:step1}
\end{equation}
In the situations of interest to us, the momentum $p$ is
relativistic, and the Lorentz violating terms are small:
\begin{eqnarray}
    m/p\ll1\\
\xi(k/M)^{n-2}\ll1\\
\eta(p/M)^{n-2}\ll1
\end{eqnarray}
Using these approximations and expanding the two energies in
powers of the small quantities $((m/p)^{2}+\eta p^{(n-2)})$ and
$\xi k^{(n-2)}$ we obtain
\begin{equation}
\frac{E_{p}}{p}\frac{\omega}{k}= \left[1+\frac{1}{2}
\left(\frac{m^{2}}{p^2}+ \eta p^{(n-2)}\right)-
\frac{1}{8}\left(\frac{m^{2}}{p^2}+ \eta p^{(n-2)}\right)^{2}
\right]\cdot \left[1+\frac{1}{2} \xi k^{(n-2)}-\frac{1}{8}
\left(\xi k^{(n-2)}\right)^{2} \right]. \label{eq:expansion}
\end{equation}
There is a subtlety about the truncation of this double expansion.
If the ratio of the two expansion parameters is very large, it is
possible that the second order term in one quantity is comparable
to (or larger than) the first order term in the other quantity. In
such cases, spurious results can be obtained by truncating both
expansions at the same order.  We shall proceed with the first
order truncation of both expansions. One can check {\it a
posteriori} whether the truncation is consistent. It turns out
that this truncation is adequate for our practical purposes. In
particular, although at very high energies our approximate
threshold results will fail to be accurate, those energies are
sufficiently high so as to be observationally irrelevant.

Another important point is that Eq.~(\ref{eq:step1}) originated
from (\ref{eq:squares}) together with conservation of
three-momentum and hence it is equivalent to energy conservation
$E(p)-\o(k)=\pm E(q)$. For the \v{C}erenkov and photon decay
processes $e^-\rightarrow e^{-}\g$ and $\g\rightarrow e^{+}e^{-}$
we want only the upper and lower signs respectively, since the
energy of all the particles should be positive. It will be
unnecessary to impose this choice explicitly however, since the
negative energy solutions are excluded by the approximations to be
employed, as can be checked by just imposing energy conservation
directly and using the same approximations. We indicate below how
the approximations can exclude the negative energy solutions.

Consider for example the vacuum \v{C}erenkov process. Then energy
conservation with a negative energy final electron reads
$E(p)=\o(k)-E(q)$ with $\o(k)$, $E(p)$ and $E(q)$ all positive.
The smallest $E(q)$ can be (within the monotonic regime) is $m$,
so we must have $\o(k)>E(p)+m$. Expanding, this becomes $k+\xi
k^{(n-1)}/2> p + m^2/2p + \eta p^{(n-1)}/2 + m$.  On the other
hand, momentum conservation (in the threshold configuration)
requires that $k<p$. This inequality implies that $\xi
k^{(n-1)}/2> \eta p^{(n-1)}/2 + m$, which requires that either
$\xi k^{(n-2)}\gtrsim O(m/p)$, or $|\eta p^{(n-2)}|\sim m/p$, or
both. In either case, we see that neglected terms such as $(\xi
k^{(n-2)})^2$ are not negligible compared to the term $(m/p)^2$
that has been kept.

Truncating (\ref{eq:expansion}) at first order, and inserting the
result in Eq.~(\ref{eq:step1}) we obtain
\begin{equation}
\xi k^n+\eta p^n-\eta |p-k|^n = 2pk\left(\frac{m^2}{2
p^2}+\frac{\xi}{2} k^{(n-2)} +\frac{\eta}{2} p^{(n-2)}\right).
\label{eq:step2}
\end{equation}
Introducing the variable $x=k/p$, (\ref{eq:step2}) takes the form
\begin{equation}
\frac{m^2}{p^n}= -\xi x^{(n-2)}\left(1-x\right)+
\eta\left[\frac{1-x-|1-x|^{n}}{x}\right]. \label{eq:step3}
\end{equation}
At threshold for either \v{C}erenkov or photon decay $p$ and $x$
must satisfy this kinematic relation.  Note that while we have
assumed that $p$ is relativistic, no such assumption is needed for
the other two momenta $q$ and $k$. This is important since we
shall use (\ref{eq:step3}) in cases where the momentum
distribution is highly asymmetric.

%----------------------------------------------------------------------
\subsection{Vacuum \v{C}erenkov effect: $e^{-}\to\gamma\,e^{-}$}
\label{sec:cerenkov}
%----------------------------------------------------------------------

The spontaneous emission of photons by a charged particle in
vacuum is forbidden in Lorentz invariant physics since the sum of
a timelike and null 4-momentum vectors cannot lie on the same mass
shell as the timelike 4-momentum. Modifications of the dispersion
relations of the form (\ref{eq:pdr}) and (\ref{eq:mdr}) can allow
some phase space for this reaction to happen. If the reaction is
allowed the rate of energy loss for the case $n>2$ well above
threshold is $dE/dt \sim\alpha E^2$, where $E$ is the energy of
the initial charged particle and $\alpha$ is the fine structure
constant.\footnote{
%-------------------------------------------------------
\label{foot:spec}For the special case $n=2$ the
rate of energy loss is further suppressed by the difference in speeds,
$dE/dt\sim (c^{2}_{e}-c^{2}_{\gamma}) \alpha E^2$, see
e.g.~\cite{CG}. In this case the decay distance depends on how close
the two speeds are, which must be taken into account in deducing
observational constraints on the parameters.}
%-------------------------------------------------------
The decay distance is thus only of order the microscopic distance
$100/E$, hence the lower threshold of the vacuum \v{C}erenkov effect
must be above the maximal observed energy of any charged particle
known to propagate.

The lower threshold is the lowest value of the incoming electron
momentum $p$ for which the kinematic equation (\ref{eq:step3}) has
a solution. At threshold $x=k/p$ must lie between 0 and 1, since
if $k$ were greater than $p$ the final electron momentum would
have to be anti-parallel to the photon momentum, which is excluded
by the threshold theorem ({\it cf.} Sect. \ref{kinematics}).  The
threshold therefore occurs at the value of $x$ in this range for
which the right hand side has a maximum. We substitute this $x$ in
Eq~(\ref{eq:step3}) to obtain the lower threshold momentum for the
electron as a function of $\xi,\eta$ and $m$. That there is no
upper threshold in this case is immediately obvious since the
right hand side vanishes as $x$ approaches one, allowing solutions
with arbitrarily large momentum.

The analysis is somewhat simplified by rewriting
Eq.~(\ref{eq:step3}) in terms of the new variable $w=1-x$, in
terms of which it takes the form
\begin{equation}
\frac{m^2}{p^n}= -\xi w(1-w)^{n-2}+ \eta(w+\cdots+w^{n-1}).
\label{eq:emce}
\end{equation}
The relevant range of $w$ is 0 to 1. In the threshold
configuration we have $p=q+k$, hence $w=q/p$. The general analysis
of the threshold relations must be done on a case by case basis
for different values of $n$, however it is easy to derive partial
results valid for any $n$.

First consider the case where $\xi$ is positive. Then the first
term in (\ref{eq:emce}) is negative, so if $\eta$ is negative
there is no solution. If $\eta$ is positive, the maximum of the
right hand side clearly occurs at $w=1$, where it is equal to
$(n-1)\eta$. Hence the threshold for the case when $\xi$ and
$\eta$ are both positive and $n>2$ is
\begin{equation}
p_{\rm th}=\left[\frac{m^2}{(n-1)\eta} \right]^{1/n}.
\end{equation}
Since $w=1$, this threshold corresponds to the emission of a zero
energy photon. This is why the value of $\xi$ is irrelevant, and
the \v{C}erenkov process takes place as long as $\eta$ is
positive. Indeed also for negative values of $\xi$ the process
takes place as long as $\eta$ is positive (and even for some
negative values---see below), however the threshold configuration
may occur with the emission of a hard photon.

One more general result can be established, namely that there is
no threshold if $\eta\le\xi\le0$. To see this observe that for $w$
between 0 and 1 we have $w+\cdots+w^{n-2}>w(1-w)^{n-2}$, since the
derivative of the lhs is greater than unity and the derivative of
the rhs is less than unity. Thus if $\eta\le\xi\le0$ the rhs of
(\ref{eq:emce}) is nowhere positive in this range of $w$. In
particular, there is no threshold in the case of equal negative
parameters $\xi=\eta<0$.

The remaining parameter space for which we need to determine the
threshold is the region $\xi<0$ and $\eta>\xi$.

%------------------------------------------------------------------%
\subsubsection{Vacuum \v{C}erenkov thresholds
for $n=2,3,4$}\label{sec:cerenres}
%------------------------------------------------------------------%

In the case $n=2$ equation (\ref{eq:emce}) becomes
\begin{equation}
\frac{m^2}{p^2}= (\eta-\xi)w, \label{eq:emce2}
\end{equation}
hence the threshold occurs at $w=1$, with the emission of a zero
energy photon, and the threshold momentum is given by
\begin{equation}
p_{\rm th}= \left(\frac{m^2}{\eta-\xi}\right)^{1/2}.
\label{eq:cer2}
\end{equation}

It is clear from the above expression that no threshold exists for
the special case $\xi=\eta$. This can also be seen directly from
the fact that quadratic modifications in the dispersion relations
are equivalent to constant (momentum independent) shifts in the
speed of propagation. In the case of equal coefficients the
electron and photon dispersion relations share the same Lorentz
symmetry only with a modified speed of light, and hence the vacuum
\v{C}erenkov effect (as well as photon decay) cannot take place.
Nevertheless we shall see that in the higher order cases ($n\geq
3$) these processes are allowed for equal positive coefficients.

In the case $n=3$ equation (\ref{eq:emce}) becomes
\begin{equation}
\frac{m^2}{p^3}= (\eta-\xi)w + (\eta+\xi)w^{2}. \label{eq:emce3}
\end{equation}
The form of the threshold relation for $p$ depends on the values
of $\eta$ and $\xi$. We find two different formulae, depending on
whether the threshold occurs with emission of a low energy photon
($w\to 1$) --- which we label as case a) below --- or with
emission of a photon with energy of order $p$ --- which is labeled
case b). After a bit of calculation we find:
\begin{eqnarray}
&{\rm a)}& \quad p_{\rm th} = \left(
\displaystyle{\frac{m^2}{2\,\eta}} \right)^{1/3} \qquad \qquad
\quad \:\:\: \mbox{for $\eta>0$ and $\xi\geq-3\eta$},
\label{cer3-a}\\
&{\rm b)}& \quad p_{\rm th} = \left[
\displaystyle{-\frac{4\,m^2\left(\xi+\eta\right)}
{\left(\xi-\eta\right)^2}} \right]^{1/3} \quad \;\mbox{for
$\xi<-3\eta<0$ or $\xi<\eta \leq 0$}
\label{cer3-b}\\
&&\nonumber\\
&{\rm c)}& \quad \mbox{No threshold}\qquad\qquad\qquad\quad
\mbox{for $\eta<0$ and $\xi>\eta$}.
\end{eqnarray}
In case b) the value of $w=q/p$ at the threshold is given by
$q/p=- (\eta-\xi)/2(\eta+\xi)$. Given a maximal energy/momentum
$p_{\rm max}$ for which no vacuum \v{C}erenkov effect is observed,
the constraint on the parameters can be written as:
\begin{eqnarray}
{\rm a)} \quad \eta&<&\displaystyle{ \frac{m^2}{2p_{\rm
max}^{3}}}, \label{eq:cerconda}
\\
{\rm b)} \quad \xi&>&\displaystyle{\eta-2\, \frac{m^2}{p_{\rm
max}^{3}}-2\sqrt{\left(\frac{m^2}{p_{\rm
max}^{3}}\right)^2-2\eta\left(\frac{m^2}{p_{\rm
max}^{3}}\right)}}. \label{eq:cercondb}
\end{eqnarray}

The case in which the correction is of quartic order is similar to
the cubic one, although somewhat more complicated. In the case
$n=4$ equation (\ref{eq:emce}) becomes
\begin{equation}
\frac{m^2}{p^4}= (\eta-\xi)w + (\eta+2\xi)w^{2}+(\eta-\xi)w^3.
\label{eq:emce4}
\end{equation}
With the definitions $\lambda\equiv\eta-\xi$ and $\tau\equiv
(\eta+2\xi)/\lambda$, (\ref{eq:emce4}) takes the form
$m^2/p^4=\lambda(w+\tau w^2+w^3)$, which is what we used to carry
out the threshold analysis. Again the form of the threshold
relation for $p$ depends on the values of $\eta$ and $\xi$, and we
label the cases with a) and b) as for $n=3$. After some tiresome
analysis we obtain the following expressions:
\begin{eqnarray}
&{\rm a)}& \quad p_{\rm th}=
\left(\frac{m^2}{3\,\eta}\right)^{1/4}
\qquad\:\:\:\, \mbox{for $\eta>0$ and $\xi\geq-(8+6\sqrt{2})\eta$},\\
&{\rm b)}& \quad p_{\rm th}=
\left(\frac{m^2}{F(\lambda,\tau)}\right)^{1/4} \quad \mbox{for
$\xi<-(8+6\sqrt{2})\eta<0$ or $\xi<\eta\leq 0$},
\label{eq:implice4}\\
&&\nonumber\\
&{\rm c)}& \quad \mbox{No threshold}\qquad\:\:\mbox{for $\eta<0$
and $\xi>\eta$},
\end{eqnarray}
where the function $F(\lambda,\tau)$ is given by
\begin{equation}
F(\lambda,\tau)=\frac{2}{27}\lambda
 \left[\tau^{3}+\left(\tau^{2}-3\right)^{3/2}-\frac{9}{2}\tau\right].
\end{equation}
In the case a) we have again the emission of a low energy photon
($w\to 1$). In case b) the value of $w=q/p$ at the threshold is
given by $q/p=(-\t-\sqrt{\t^2-3})/3$.  So for $n=4$ given a
maximal energy/momentum (say $p_{\rm max}$) for which no vacuum
\v{C}erenkov effect is observed, the constraint for case a) can be
written as:
\begin{equation}
{\rm a)} \quad \eta<\displaystyle{ \frac{m^2}{3p_{\rm max}^{4}}}.
\end{equation}
The constraint for case b) has a cumbersome form but the
corresponding line in the $\xi$--$\eta$ plane can be found from
equation (\ref{eq:implice4}).

%--------------------------------------------------------%
\subsubsection{Observations and constraints from
absence of vacuum \v{C}erenkov effect} \label{sec:obcerenkov}
%--------------------------------------------------------%

We can now consider the actual constraints observations impose on
$\xi$ and $\eta$. The previous analysis shows that the smallness
of $m^2/p_{max}^n$ determines the strength of the constraint
provided by the vacuum \v{C}erenkov effect, hence the strongest
constraint will be obtained by considering the highest energy
observed for a given particle.

Electrons in particle accelerators are stable against the
vacuum \v{C}erenkov effect at energies up to $500$ GeV,
and in cosmic rays energies of $\sim 2$ TeV have been
detected~\cite{Kifune:1999ex,nishi}.  Even higher
energies, in the range $50-100$ TeV, are necessary in
order to consistently explain the peaks in the X-ray and
TeV regions of the photon emission from supernova
remnants such as SNR1006 or the Crab
Nebula~\cite{Koyama,AAK,Kifune:1999ex}.  In particular
for the Supernova remnant SN1006 a clear identification
of a synchrotron emission together with the independent
estimate of the magnetic field strength allows one to
infer that the electrons should have energies of about
100 TeV~\cite{Koyama,Naito}.\footnote{After this work
  was completed we found~\cite{Crab} that the synchrotron
  emission is sensitive to Lorentz violation, and in fact
  one cannot be certain about the existence of these 100
  TeV electrons for positive $\eta$. However one can
  instead use the existence of 50 TeV electrons inferred
  from the detection of 50 TeV photons produced by
  inverse Compton scattering in the Crab nebula. This
  would weaken the constraint by just a factor of $2^{3}=8$.
}. These electrons propagate over distances far longer
than that required by the vacuum \v{C}erenkov effect to
decrease the electron energy below the
threshold.\footnote{
%------------------------------------------------------------------
The competing energy loss by synchrotron radiation is irrelevant
for this constraint. The rate of energy loss from a particle of
energy $E$ due to the vacuum \v{C}erenkov effect goes like $-e^2
E^2$, while that from synchrotron emission goes like $- e^4 B^2
E^2/m^4$ (using units where $c=\hbar=1$). For a magnetic field of
about one micro Gauss (such as those involved in supernova
remnants) the synchrotron emission rate is $40$ orders of
magnitude smaller than the vacuum \v{C}erenkov rate. }
%----------------------------------------------------------

\begin{figure}[htb]
\vbox{ \vskip 8 pt
\centerline{\includegraphics[width=2.7in]{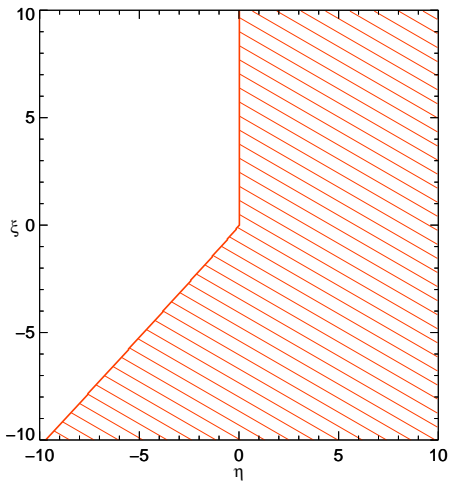}}
%\centerline{\epsfxsize=3.25in\epsffile{cer-n3-ph.eps}}
\caption{\label{fig:cer-n3-ph} Constraint from the absence of
vacuum \v{C}erenkov effect for $n=3$. The filled region in the
parameter space is the one not compatible with the existence of
the $\sim 100$ TeV electrons indirectly detected via synchrotron
emission from supernova remnants~\cite{Koyama}. The point where
the vertical line crosses the $\eta$ axis is $\eta=m^2/(2p^3_{\rm
max})\sim 1.5\times10^{-3}$.
\smallskip} }
\end{figure}

For $n=2$ we see from Eq.~(\ref{eq:cer2}) that $(\eta-\xi)\lesssim
3\times 10^{-17}$ which can be compared with the limit
$(\eta-\xi)\lesssim 5\times 10^{-13}$ obtained by Coleman and
Glashow~\cite{CG} using a $p_{max}$ of 500 GeV.  The \v{C}erenkov
emission rate (cf footnote~\ref{foot:spec}) is fast enough for
such parameters that $\Delta E\sim E$ over a distance scale of
centimeters. For $n\geq 3$ the emission rate is $10^{17}$ times
higher.  For the cases of $n=3$ and $n=4$ the corresponding value
of $m^2/p^n_{max}$ is $\sim 3\times 10^{-3}$ and $\sim 4\times
10^{11}$ respectively. We therefore obtain an interesting
constraint for the cubic case but not for the quartic case,
assuming that the Lorentz violation is at the Planck scale. (We
shall see in section~\ref{sec:alt-cer-p} that one could get a good
constraint even for the $n=4$ case by considering the $10^{20}$ eV
cosmic ray protons, modulo some caveats that we shall discuss.)
Figure~\ref{fig:cer-n3-ph} shows the excluded region for the
parameters $\xi$ and $\eta$ in the $n=3$ case as determined by the
conditions (\ref{eq:cerconda}) and (\ref{eq:cercondb}).
%------------------------------------------------------------------
%---------------------------------------------------------------------

%%%%%%%%%%%%%%%%%%%%%%%%%%%%%%%%%%%%%%%%%%%%%%%%%%%%%%%%%%%%
\subsection{Photon Decay: $\gamma\to e^{+}\,e^{-}$}
\label{sec:gdecay}
%%%%%%%%%%%%%%%%%%%%%%%%%%%%%%%%%%%%%%%%%%%%%%%%%%%%%%%%%%%%

The spontaneous decay of a photon into a electron-positron pair,
is another reaction usually forbidden by energy--momentum
conservation. As in the case of the vacuum \v{C}erenkov effect,
modifications of the dispersion relation allow this reaction to
occur. By the threshold theorem ({\it cf}.
Sect.~\ref{kinematics}), we know that the final particles have
parallel momenta, so that both lepton momenta are less than or
equal to $k$. Thus $x:=k/p\ge1$ in Eq. (\ref{eq:step3}), so that
$|1-x|=x-1$. It is convenient to use the variable $y:=1/x=p/k$,
whose relevant range is zero to one. In terms of y,
(\ref{eq:step3}) takes the form
\begin{equation}
\frac{m^{2}}{k^{n}}=\xi y \left(1-y\right)-\eta y \left(1-y\right)
\left[ y^{(n-1)}+\left(1-y\right)^{(n-1)}\right].
    \label{eq:gdec}
\end{equation}
The threshold corresponds to the maximum of the right hand side of
Eq.\ (\ref{eq:gdec}) with respect to $y$. Note that the rhs is
symmetric about $y=1/2$, since the two leptons are kinematically
interchangeable, hence it is always stationary at $y=1/2$.
However, this stationary point can be a maximum or a minimum,
depending on the values of $\eta$ and $\xi$. If it is a maximum
the threshold momentum is given by
\begin{equation}
k_{\rm th}=\left[ \frac{2^{n} m^{2}}{2^{(n-2)}\xi-\eta}
\right]^{1/n}.
\end{equation}

In the special case $\xi=\eta$, which has been mostly studied in
the literature, it can be shown that the only stationary points of
(\ref{eq:gdec}) are $y=0, 1/2, 1$. Given that the right hand side
of Eq.~(\ref{eq:gdec}) is always zero at $y=0,1$ it follows that
for equal coefficients the threshold condition is always realized
with a symmetric distribution of the final momenta.

Contrary to relativistic intuition, and to what has been assumed
in all previous calculations as far as we know, the threshold does
{\it not} always occur with the symmetric configuration. The
reason is that when $\eta<0$, the lepton energy $E(p)$ has
negative curvature $E''(p)<0$ for sufficiently large momentum if
$n>2$, unlike the usual Lorentz invariant case. If the threshold
lies within the negative curvature region, it cannot occur with
the symmetric configuration since the energy of the final state at
fixed momentum could be lowered by making the momentum of one
particle smaller and one larger by an equal amount. For
$\eta<\xi<0$, the threshold does occur in the negative curvature
region, hence it is asymmetric.

The occurrence of the asymmetric threshold might seem especially
surprising if we think, with relativistic habits, that at
threshold the electron and positron should be created at rest in
the center of mass frame. The error lies in a misleading
application of the Lorentz transformation in the case where a
definite preferred system exists. First, the center of mass frame
may not even be accessible if the photon energy-momentum vector is
spacelike (i.e. subluminal dispersion). Second, even if we can
boost to the center of mass frame, in this frame the dispersion
relation of the electron/positron may not have its minimum energy
at zero momentum. Therefore it is not always true that the final
particles are produced at rest in the center of mass frame.

We now examine the cases $n=2,3,4$ individually.

%---------------------------------------------------%
\subsubsection{Photon decay thresholds for $n=2,3,4$}\label{gdecayres}
%----------------------------------------------------%

In the case $n=2$ Eq. (\ref{eq:gdec}) takes the form
\begin{equation}
\frac{m^{2}}{k^{2}}=(\xi-\eta)y(1-y).
    \label{eq:gdec2}
\end{equation}
For $\xi-\eta<0$ there is no threshold, while for $\xi-\eta>0$
there is a lower threshold at $y=1/2$. In this case one obtains
the threshold formula
\begin{equation}
k_{\rm th}=\frac{2 m} {\sqrt{\xi-\eta}}
    \label{eq:gdec2th}
\end{equation}

%--------------------------%
%\subsubsection{n=3}
%--------------------------%

In the case $n=3$ (\ref{eq:gdec}) takes the form
\begin{equation}
 \frac{m^{2}}{k^{3}}=\xi y \left(1-y\right)-\eta y \left(1-y\right)
  \left[ y^{2}+\left(1-y\right)^{2}\right],
   \label{eq:gdec3}
\end{equation}
To determine the threshold we need to find the maximal values of
the rhs. The task of finding the maxima is simplified by
introducing the new variable $z=(2y-1)^2$, so that
$y=(1+\sqrt{z})/2$, $(1-y)= (1-\sqrt{z})/2$, and $y(1-y)=
(1-z)/4$. The relevant range of $z$ is $[0,1]$, where $z=0$
corresponds to the symmetric configuration $y=1/2$ and $z=1$
corresponds to $y=1$.

In terms of $z$, (\ref{eq:gdec3}) becomes
\begin{equation}
 \frac{m^{2}}{k^{3}}=\frac{\xi}{4}(1-z)-\frac{\eta}{8} (1-z^2)
   \label{eq:gdec3z}
\end{equation}
The symmetric extremum at $y=1/2$ corresponds to $z=0$, and there
is one other (asymmetric) extremum at $z_a=\xi/\eta$. One of the
two extrema is a maximum and the other is a minimum. Since the
second derivative with respect to $z$ is $\eta/4$, the one at
$z_a$ is a maximum~\footnote{
%----------------------------------------
This does not also show that the extremum at $z=0$ is a maximum,
since the relation between $z$ and $y$ is not smooth there. In
fact, $d^2\!/dy^2 = 16 z\, d^{2}\!/dz^{2} + 8\, d/dz$, so at $z=0$ we have
$d^2\!/dy^2 = 8\,d/dz$. Using this we see that the symmetric solution
is a maximum if and only if $\xi>0$, so the asymmetric solution is
the maximum if and only if $\xi<0$. This is the same condition as
$\eta<0$, since if $z_a=\xi/\eta$ is greater than zero, $\xi<0$ if
and only if $\eta<0$.
%----------------------------------------
} if and only if $\eta<0$, and it lies between zero and one in
this case if and only if $\eta<\xi<0$. Note that in the special
case $\xi=\eta$ the asymmetric threshold solution is removed and a
threshold exists just for positive values of $\eta$.

The value of the rhs of (\ref{eq:gdec3z}) at $z=0$ is
$(2\xi-\eta)/8$, while at $z=z_a$ it is $-(\xi-\eta)^2/8\eta$. We
thus see that photon decay is allowed only above the broken line
in the $\eta$--$\xi$ plane given by $\xi=\eta/2$ in the quadrant
$\xi,\eta>0$ and by $\xi=\eta$ in the quadrant $\xi,\eta<0$. Above
this line, the threshold is given by
\begin{eqnarray}
 \mbox{a)}\quad
   k_{\rm th}&=& \displaystyle{ \left( \frac{8 m^2}{2\xi-\eta} \right)^{1/3}}
   \quad \mbox{for $\xi\geq0$},\label{th1}\\
 \mbox{b)}\quad
   k_{\rm th}&=&
\displaystyle{\left[\frac{-8 m^2\eta}{\left(\xi-\eta \right)^2}
\right]^{1/3}}
   \quad \mbox{for $\eta<\xi<0$}.\label{th2}
\end{eqnarray}

The detection of gamma rays with momenta up to some $k_{\rm max}$
implies that the parameters must lie in the $\xi-\eta$ plane below
the line corresponding to a threshold at $k_{\rm max}$. This
translates into the following constraints for the parameters $\xi$
and $\eta$
\begin{equation}
\mbox{a)} \quad
\xi<\displaystyle{\frac{\eta}{2}+\frac{4m^2}{k^{3}_{\rm
max}}},\qquad \mbox{b)} \quad \xi<
\displaystyle{\eta+\sqrt{-\frac{8m^2\eta}{k^{3}_{\rm max}}}}.
\end{equation}

In the case $n=4$ Eq.~(\ref{eq:gdec}) can again be conveniently
rewritten in terms of the variable $z$ introduced after
(\ref{eq:gdec3}) above, yielding
\begin{equation}
 \frac{m^{2}}{k^{4}}=\frac{\xi}{4}(1-z) -\frac{\eta}{16}(1+2z-3z^2).
   \label{eq:gdec4z}
\end{equation}
The asymmetric extremum here occurs at $z_a=(2\xi+\eta)/3\eta$.
This is again a maximum if and only if $\eta<0$, and it lies
between zero and one in this case if and only if
$\eta<\xi<-\eta/2$. Note that again in the special case $\xi=\eta$
the asymmetric threshold solution is removed and a threshold
exists just for positive values of $\eta$.

The value of the rhs of of (\ref{eq:gdec4z}) at $z=0$ is
$(4\xi-\eta)/16$, while at $z=z_a$ it is $-(\xi-\eta)^2/12\eta$.
We thus see that photon decay is allowed only above the broken
line in the $\eta$--$\xi$ plane given by $\xi=\eta/4$ in the
quadrant $\xi,\eta>0$ and by $\xi=\eta$ in the $\xi,\eta<0$. Above
this line, the threshold is given by
\begin{eqnarray}
 \mbox{a)}\quad
   k_{\rm th}&=& \displaystyle{\left( \frac{16 m^2}{4\xi-\eta}
   \right)^{1/4} }
   \quad \mbox{for $\xi\geq-\eta/2$},\label{th1-4}\\
 \mbox{b)}\quad
   k_{\rm th}&=&
\displaystyle{ \left[ \frac{-12 m^2 \eta}{\left(\xi-\eta\right)^2}
\right]^{1/4} }
   \quad \mbox{for $\eta<\xi<-\eta/2$}.\label{th2-4}
\end{eqnarray}

Again, given a maximal observed momentum for which gamma decay is
not observed gives constraints on the parameters $\xi$ and $\eta$
\begin{equation}
\mbox{a)} \quad \xi< \displaystyle{ \frac{\eta}{4} +\frac{4 m^2}{
k^{4}_{\rm max} } }, \qquad \mbox{b)} \quad \xi< \displaystyle{
\eta+\sqrt{ -\frac{12 m^2 \eta}{k^{4}_{\rm max} }} }.
\end{equation}
%

%--------------------------------------------------------------
\subsubsection{Observations and constraints from absence of
photon decay}
%--------------------------------------------------------------

We can now consider the constraint on $\xi$ and $\eta$ imposed by
the absence of photon decay in current observations. As before,
the smallness of $m^2/k_{max}^n$ determines the strength of the
constraint, hence the strongest constraint will be obtained by
considering the highest energy photons observed, which are the 50
TeV gamma rays arriving on earth from the Crab
nebula~\cite{Tanimori}. The rapid decay rate ($\Gamma \sim E$
above threshold) implies that in order to propagate at all, let
alone to reach us from the Crab nebula, these photons must have an
energy below the threshold. For the 50 TeV photons we have
$m^2/k_{max}^n\sim 10^{14n-44}$. For $n=2$ and $n=3$ this yields
strong constraints on $\xi$ and $\eta$, however for $n\ge4$ this
number is $\sim 10^{12}$ so the constraints are not so
interesting. The case $n=2$ has already been studied
in~\cite{CG,SG01}. Reference~\cite{SG01} also uses 50 TeV, which
from Eq.~(\ref{eq:gdec2th}) yields the constraint
$(\xi-\eta)\lesssim 10^{-16}$. Here we consider the case $n=3$.

\begin{figure}[htb]
\vbox{ \vskip 8 pt
\centerline{\includegraphics[width=2.7in]{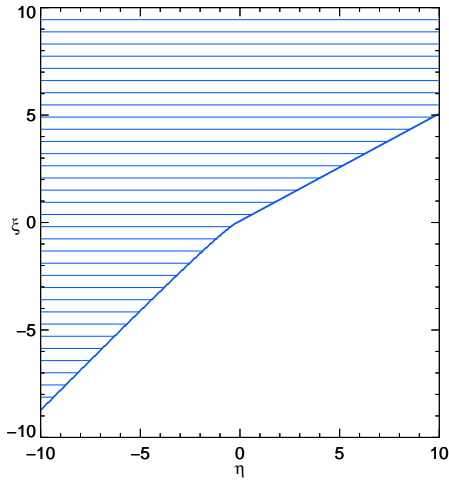}}
%\centerline{\epsfxsize=3.25in\epsffile{gdec-n3-ph.eps}}
\caption{Constraint from the absence of photon decay. The filled
region in the parameter space is the one excluded by the
observation of gamma rays of energies up to $\sim 50$
TeV.\label{fig:decay}
\smallskip}
}
\end{figure}

In the case $n=3$, we use expressions (\ref{th1}) and (\ref{th2})
for the threshold momenta to impose the condition that photon
decay be forbidden for photons below $k_{\rm max}=50$ TeV. This
defines a broken line in the $\xi$--$\eta$ plane below which the
coefficients must lie:
\begin{equation}
\mbox{a)} \quad \xi<\displaystyle{\frac{\eta}{2}+0.08},\qquad
\mbox{b)} \quad \xi< \displaystyle{\eta+\sqrt{-0.16\,\eta}}.
\end{equation}
Constraint (a) applies for $\xi>0$ while (b) applies for $\xi<0$.
The excluded region in the parameter space is is shown in
Fig.~\ref{fig:decay}.
%--------------------------------------------------------------------------%
%---------------------------------------------------------------------------%

\begin{figure}[htb]
\vbox{ \vskip 8 pt
\centerline{\includegraphics[width=2.3in]{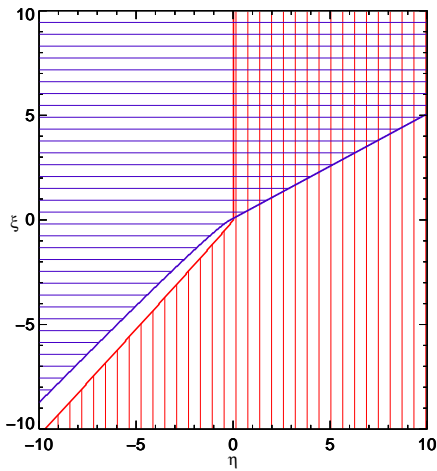}} \caption{The
graph shows the combined observational constraint derived from the
absence of vacuum \v{C}erenkov effect and gamma decay. The
horizontal blue shading identifies the region excluded by gamma
decay, the vertical red one by \v{C}erenkov. Although not visible
there is a tiny region of positive $\xi$ and $\eta$ allowed by
present observations, and there is a barely visible region of
positive $\xi$ and negative $\eta$. Also the diagonal is in the
interior of the allowed region. \label{fig:cergdec}
\smallskip}
}
\end{figure}

The joint constraints imposed by both vacuum \v{C}erenkov and
photon decay are shown in Fig.~\ref{fig:cergdec}. We see that
these two reactions are already enough for ruling out most of the
parameter space. Next we shall see that by taking into account
also the process of photon annihilation this constraint can be
further improved.
%
%--------------------------------------------------------------------------%
%---------------------------------------------------------------------------%

%----------------------------------------------------------------------------%
\subsection{Photon annihilation: $\gamma\,\gamma \to e^{+}\,e^{-}$}
\label{sec:gg}
%----------------------------------------------------------------------------%

In standard QED two photons can annihilate to form an
electron-positron pair. If one of the photons has energy $\omega_0$,
the threshold for the reaction occurs in a head-on collision with the
second photon having the momentum (equivalently energy) $k_{\rm
LI}=m^2/\omega_{0}$. For $k_{\rm LI}= 10$ TeV (which will be relevant
for the observational constraints) the soft photon threshold
$\omega_0$ is approximately 25 meV, corresponding to a wavelength of
50 microns.

In the presence of Lorentz violating dispersion relations the
threshold for this process is in general altered, and the process can
even be forbidden. Moreover, as noticed by Klu\'zniak~\cite{Kluzniak},
in some cases there is an upper threshold beyond which the process
does not occur.\footnote{
%-------------------------------------------------------
As discussed below in section \ref{subsec:uppersummary}, our results
agree with those of \cite{Kluzniak} only in certain limiting cases.}
%-------------------------------------------------------
In this section we discuss how the thresholds depend on the Lorentz
violating parameters. We then discuss the observational consequences
and constraints that can be obtained using the absorption of TeV gamma
rays of extragalactic origin by the intervening infrared (IR)
background.

The threshold equation for photon annihilation can be obtained by
modifying our previous analysis of photon decay. The difference is
that the initial state includes two photons rather than one. We are
interested in the case where one of the photons has low energy (IR),
hence for that photon the modification in the dispersion relation can
be neglected. The threshold theorem ({\it cf}.
Sect.~\ref{kinematics}) tells us that the threshold configuration is a
head-on collision. Denoting the IR photon energy by $\omega_0$, the
total four-momentum of the initial state thus takes the form $k_{\rm
4, in}=(\o(k) + \o_0,k-\o_0,0,0)=:(\o',k',0,0)$.

To adapt our previous calculation, we need only replace $k$ by
$k'=k-\o_0$ and $\o(k)$ by $\o'(k') =\o(k'+\o_0) + \o_0$.  Expanding
one gets $\o(k'+\o_0)= k'+\o_0 + (\xi/2)(k'+\o_0)^{n-1}+\cdots$. Since
$\o_0\ll k$, and the last term is already Planck-suppressed (or, if
$n=2$, suppressed by the small value of $\xi$), we can neglect $\o_0$
in that term. This yields the approximation $\o'(k')=k' +
(\xi'/2)(k')^{(n-1)}$, where $\xi'$ is defined by
\begin{equation}
 \xi^{'}=\xi+\frac{4\omega_{0}}{(k')^{(n-1)}}.
  \label{eq:shift}
\end{equation}
The kinematic equation for photon annihilation is thus obtained from
that for photon decay (\ref{eq:gdec}) by the replacements
$k\rightarrow k'$ on the lhs and $\xi\rightarrow \xi'$ on the rhs.  We
can further neglect the difference between $k'$ and $k$ on the lhs
since $\o_0\ll k$, hence to a sufficiently good approximation we can
use the kinematic equation
\begin{equation}
 0=F(k,y):=-\frac{m^{2}}{k^{n}}+\left(
   \xi+\frac{4\omega_{0}}{k^{(n-1)}} \right) y \left(1-y\right)-\eta y
   \left(1-y\right) \left[ y^{(n-1)}+\left(1-y\right)^{(n-1)}\right].
   \label{eq:ggscat}
\end{equation}
The variable $y$ is defined by $y=p/k$, where $p$ is one of the
lepton momenta. Our analysis of the thresholds is based on Eq.
(\ref{eq:ggscat}).

As in the case of photon decay, the thresholds occur at the symmetric
value $y=1/2$ only for certain ranges of the parameters $\xi$ and
$\eta$. The analysis for photon annihilation is more complicated
however since for $n\ge3$ the dependence of Eq.  (\ref{eq:ggscat}) on
$k$ and $y$ does not separate, unlike in Eq.(\ref{eq:gdec}). Thus it
is not simply a matter of finding the value of $y$ between zero and
unity for which $F(k,y)$ is maximum.  Analyzing the threshold
structure is a rather lengthy and complicated process, so we have
placed the details in an Appendix.  The analysis reveals a number of
unexpected features that thresholds can have in the presence of
Lorentz violating dispersion, with intricate dependence on the Lorentz
violating parameters. Here we summarize the results in the cases
$n=2,3$, and apply them to obtain further observational constraints.

We obtain results valid for any value of the soft photon energy
$\omega_0$ and ``electron'' mass $m$ by employing appropriately
scaled quantities:
\begin{equation}
\b=k/k_{\rm LI},\qquad \widetilde{\eta}= \eta
(m^{2(n-1)}/\omega_0^n),\qquad \widetilde{\xi}=\xi
(m^{2(n-1)}/\omega_0^n) \label{eq:scaled}
\end{equation}
where $k_{\rm LI}$ is the standard lower threshold $m^2/\omega_0$ The
basic threshold structure will be given in terms of these
variables. For the case of most interest to us, $n$ is 3 and $m$ is
the electron mass. For $\omega_0=25$ meV we then have
$\xi=(\o_0^3/m^4)\, \widetilde{\xi}\simeq 2.3\,\widetilde{\xi}$, and
similarly for $\eta$.

It is worth noting that while we have been thinking of $\o_0$ as fixed
and determining the corresponding high energy threshold, it can be
viewed the other way around.  The parameter $\b$ can also be written
as $\omega_0/(m^2/k)$.  If now $k$ is considered fixed then $\omega_0$
is the modified soft photon threshold and $m^2/k=\omega_{\rm LI}$ is
the corresponding Lorentz invariant threshold. Therefore $\b$ has also
the interpretation $\o_0/\o_{\rm LI}$, that is the factor by which the
soft photon threshold is shifted at fixed hard photon energy $k$. This
interpretation is valid for lower thresholds only however. There is in
fact {\it never} an upper threshold for the soft photon at fixed
$k$ (as long as $\o_0\ll k$).

%--------------------------------------------------------%
\subsubsection{Photon annihilation thresholds for n=2}
%--------------------------------------------------------%

For $n=2$ the threshold configuration is always the symmetric one.
The contour of threshold $\b$ is given by the straight line
\begin{equation}
\widetilde{\xi}=\widetilde{\eta} + 4 \frac {1-\b} {\b^2}.
\label{n2contour}\end{equation}
The $\widetilde{\xi}$-intercept decreases monotonically from $\infty$
to $-1$ for $\b<2$, and increases monotonically from $-1$ to $0$ for
$\b>2$. Hence the process is forbidden below the line
$\widetilde{\xi}=\widetilde{\eta}-1$.  The parameter $\b$ gives the
lower threshold for $\b<2$ and the upper threshold for $\b>2$.  If the
lower threshold is greater than unity, then the upper threshold exists
and is given by $\b/(\b-1)$.  The maximum lower threshold $\b=2$
corresponds to $k = 2k_{\rm LI}= 2m^2/\omega_0$.

%----------------------------------------------------------------%
\subsubsection{Lower threshold of photon annihilation for $n=3$}
%----------------------------------------------------------------%

\begin{figure}[htb]
\vbox{ \vskip 8 pt
\centerline{\includegraphics[width=2.7in]{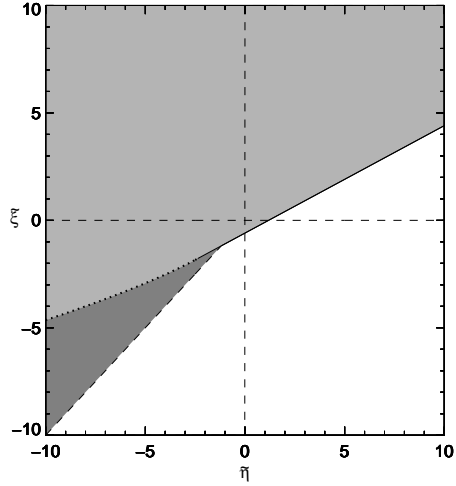}}
\caption{\label{fig:thstruct} Regions where the lower threshold
for photon annihilation with $n=3$ is determined by the symmetric
configuration (light grey region), the asymmetric one (dark grey
region) or the reaction does not occur (white region). The dotted
line is the locus of points where the contour of constant
$\beta\leq1.5$ switches smoothly from the asymmetric to the
symmetric solution.
\smallskip}
}
\end{figure}

\begin{figure}[htb]
\vbox{ \vskip 8 pt
\centerline{\includegraphics[width=2.7in]{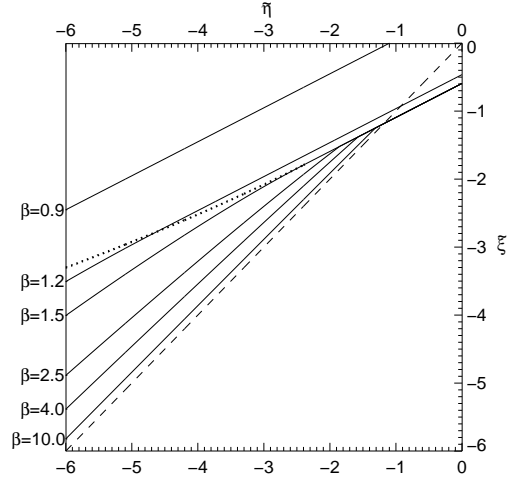}}
\caption{\label{fig:ggstruct} Contours of constant lower threshold
$\b={\rm const}$. For $\b\leq 1.5$ these correspond to symmetric
configurations above the dotted line and asymmetric ones below.  For
$\b>1.5$ there are only asymmetric lower thresholds and the contours
terminate at the symmetric $\b=1.5$ contour. The end of the dotted
line is at $(\widetilde{\eta},\widetilde{\xi})=(-64/27,\, -48/27)$,
and the $\b=1.5$ contour meets the diagonal at
$\widetilde{\eta}=\widetilde{\xi}=-32/27$.
\smallskip}
}
\end{figure}

For $n=3$ the threshold configuration is not always symmetric in the
outgoing momenta. Instead of straight parallel lines for the contours
of threshold $\b$ we find a more complicated structure.  Figure
\ref{fig:thstruct} shows the regions in the parameter plane where the
threshold configuration is symmetric or asymmetric or does not exist
at all, and a contour plot of the lower threshold is shown in Figure
\ref{fig:ggstruct}.
%---------------------------------------------------------------------------%
%---------------------------------------------------------------------------%
%---------------------------------------------------------------------------%
%---------------------------------------------------------------------------%

The threshold can be symmetric only for $\b\le1.5$. The symmetric
part of the contour is given by the straight line
\begin{equation}
\widetilde{\xi}=\frac{\widetilde{\eta}} {2} + \frac {4
(1-\b)}{\b^3}, \label{symmn3}
\end{equation}
restricted to the region above the line $\widetilde{\xi}=-4/\b^2$.
Below this line the $\b$-contour switches to the asymmetric
threshold, and is given by
\begin{equation}
\widetilde{\xi}=\widetilde{\eta} - \frac {4} {\b^2} +
\sqrt{-\frac{8\widetilde{\eta}}{\b^3}}. \label{asymmn3}
\end{equation}
The joining point of the symmetric and asymmetric parts of the
$\b$-contour is at $(\widetilde{\eta}_{\rm join},\widetilde{\xi}_{\rm
join})=(-8/\b^3,\, -4/\b^2)$.  As $\b$ varies from 0 to $1.5$ these
joining points trace out the curve $\widetilde{\xi}_{\rm join}=-
(-\widetilde{\eta})^{2/3}$.  The asymmetric threshold contours for
$\b>1.5$ terminate at the symmetric $\b=1.5$ contour, and accumulate
above the diagonal as $\b\rightarrow\infty$.  The precise degree of
asymmetry at threshold, i.e. the ratio of electron momentum to
incoming hard photon momentum, is given by $y=(1\pm\sqrt{z_a})/2$,
where $z_a=(\widetilde{\xi} + 4/{\b^2})/\widetilde{\eta}$.

%----------------------------------------------------------------%
\subsubsection{Upper threshold of photon annihilation for $n=3$}
\label{subsec:uppersummary}
%----------------------------------------------------------------%

\begin{figure}[htb]
\vbox{ \vskip 8 pt
\centerline{\includegraphics[width=2.7in]{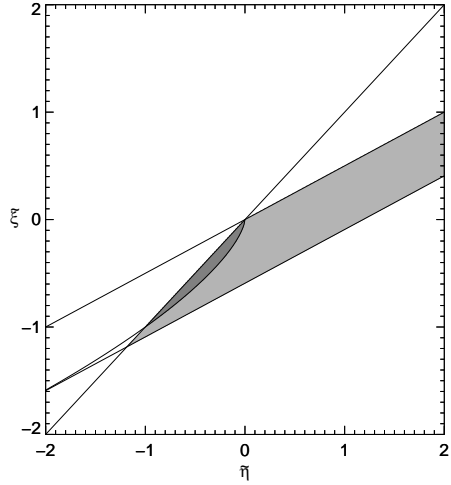}}
\caption{\label{fig:upperregions} Regions where the upper
threshold is determined by the symmetric configuration (light grey
region) or the asymmetric one (dark grey region). In the white
region below the light grey and below the diagonal the reaction
never occurs, and in the rest of the white region there is a lower
threshold but no upper threshold.
\smallskip}
}
\end{figure}

Upper thresholds exist for $n=3$ only below the diagonal and between
the $\b=1.5$ and $\b=\infty$ (which gives the same line as $\b=1$)
symmetric contours (\ref{symmn3}). For a given $\b$ the threshold is
symmetric in the region above the line $\widetilde{\xi}=-4/\b^2$ and
asymmetric below, where the contour is given by the curve
(\ref{asymmn3}). The regions of symmetric and asymmetric upper
thresholds for $n=3$ are shown in Figure~\ref{fig:upperregions}.
%------------------------------------------------------------------
%------------------------------------------------------------------
The boundary of the lens shaped region next to the diagonal is
determined by the curve $\widetilde{\xi}_{\rm join}=-
(-\widetilde{\eta})^{2/3}$ consisting of the points where the
symmetric and asymmetric segments join. The bottom of the lens meets
the diagonal at $\widetilde{\eta}=\widetilde{\xi}=-1$ where the
symmetric $\b=2$ line crosses, so asymmetric upper thresholds exist
only for $\b>2$. The lower boundary of the region of upper thresholds
is the $\b=1.5$ line, which meets the diagonal at
$\widetilde{\eta}=\widetilde{\xi}=-32/27$.

The possibility of upper thresholds for photon annihilation has been
previously discussed by Klu\'zniak~\cite{Kluzniak}, who gave results
for the values $\eta=0$, $\xi=-1$, and $\eta=\xi=-1$ in the $n=3$
case. It seems that only the symmetric configuration was examined in
\cite{Kluzniak}, hence his results cannot fully agree with ours in
cases where the asymmetric configuration is important. For the case
$\eta=0$, and negative $\xi$, our results show that there is a
symmetric upper threshold only for $\widetilde{\xi}$ values above the
$\beta=1.5$ line, i.e. for $\widetilde{\xi}>-16/27$. Our upper
threshold agrees with that of \cite{Kluzniak} in the limit
$|\widetilde{\xi}/4|=|\xi m^4/4\omega_0^3|\ll1$. The left hand side is
unity for $\xi=-1$ and $\omega_0\simeq 20$ meV, hence our results
agree approximately provided $\omega_0$ is greater than about $\simeq
40$ meV. In the diagonal case, while our results for the symmetric
configuration agree in the same limit, we have seen that there is no
upper threshold since asymmetric configurations exist for arbitrarily
large $\b$.

%------------------------------------------------------------------
\subsubsection{Observations and constraints from absence of deviations
from standard photon annihilation}\label{sec:photann}
%------------------------------------------------------------------

The \v{C}erenkov and photon decay constraints leave open an infinite
wedge-shaped region including the diagonal in the lower left quadrant
for the case $n=3$.  A constraint from agreement with standard photon
annihilation would be complementary to these and hence has the
potential to confine the allowed region to a small neighborhood of the
origin. Such a constraint is provided by indirect observations of
annihilation of high energy gamma rays from blazars on the cosmic
background radiation (CBR). Since there is presently considerable
uncertainty regarding both the background radiation and the nature of
the sources, the constraint that can be extracted is not yet very
precise however.

Another limitation of the present work arises from the fact that each
observed gamma ray has the opportunity to interact with soft photons
at any energy above the threshold, so to compare with observation one
should compute the absorption using the Lorentz violating dispersion
relation, integrating over all target frequencies.  Such an
investigation lies outside the scope of the present paper, so we shall
only attempt to roughly characterize how large a threshold shift might
be compatible with current observations.

We now summarize the observational situation. The BL Lac
objects Mkn 421 and Mkn 501 are a type of blazar emitting
high energy gamma rays whose observed spectrum reaches 17
TeV in the case of Mkn 421~\cite{Mkn421} and 24 TeV in
the case of Mkn 501~\cite{Mkn501}. The source power
spectra are reconstructed accounting for absorption via
photon annihilation on the intervening CBR, which ranges
from the near infrared (NIR, $\sim 1 ~\mu$m) to the
cosmic microwave background (CMBR, $\sim 1000
~\mu$m). Currently we have a good knowledge of the NIR
and CMBR but uncertainties remain regarding the
distribution in the intermediate, mid infrared ($\sim 10
~\mu$m) and far infrared ($100 ~\mu$m), regions (see
e.g. Figure 1 of~\cite{Aharonian:2001cp} or the
discussion in~\cite{stecker01}).  Some models of the IR
background imply a source spectrum for Mkn 501 with an
unexpected amount of radiation (a ``pile-up'') above $10$
TeV~\cite{Protheroe:2000hp,Aharonian:2001cp}. If such IR
backgrounds are correct, the pile-up might be due to a
process producing enhanced emission at energies larger
than $10$ TeV~\cite{Aharonian:2001cp}, or it might be
explained by anomalously low absorption caused by an
upward shift of the threshold due to Lorentz
violation~\cite{ACP,
  Kifune:1999ex,Kluzniak,Protheroe:2000hp,Aloisio:2000cm}.
However, recent work~\cite{SG01,stecker01} based on
improved reconstructions of the FIRB and on a new
analysis of the gamma ray flux from Mkn 501 supports the
view that current observations are consistent with the
predictions of standard Lorentz invariant theory up to 20
TeV.  Even without resolving the question of the pile-up,
it seems well established that some degree of photon
absorption has been observed up to 20 TeV, which already
provides an interesting constraint on Lorentz
violation. Moreover, it is our impression that the
suggestions of an anomaly above 10 TeV will likely prove
illusory as new observations are made available,
confirming the results
of~\cite{SG01,stecker01}~\footnote{After this work was
  completed a further observational analysis
  appeared~\cite{Konopelko:2003zr}. This allows
  the observational basis for the constraint discussed
  in this paper to be solidified~\cite{GACcom}.}.
We can thus obtain observational constraints from the requirement that
the Lorentz violation does not too strongly modify standard
Lorentz-invariant thresholds for photon annihilation. The strength of
the constraints depends of course on the order $n$ of the Lorentz
deformation. The general threshold equation (\ref{eq:gfggsapp}) shows
that an order unity constraint on $\b$ translates into an order unity
constraint on $\widetilde{\eta}$ and $\widetilde{\xi}$, which
corresponds to an order $\o_0^n/m^{2(n-1)}$ constraint on $\eta$ and
$\xi$. Since all studies seem to agree that more or less standard
Lorentz-invariant absorption is occurring for gamma rays up to 10 TeV,
we shall use the corresponding soft photon threshold of $\o_0=25$ meV
$\sim$ 50 $\m$m as a numerical benchmark. One then has $\o_0^2/m^2\sim
10^{-15}$ for $n=2$, $\o_0^3/m^4\sim 1$ for $n=3$, and $\o_0^4/m^6\sim
10^{15}$ for $n=4$. Hence only the $n=2$ and $n=3$ cases can provide
interesting constraints. Note that in the $n=3$ case, which is of most
interest to us, the dependence on $\o_0$ is cubic, so for example a
constraint at $2\, \o_0$ is eight times weaker than a constraint at
$\o_0$, while one at $\o_0/2$ is eight times stronger. This means also
that there could be strong deviations in absorption for, say, 20 TeV
gamma rays, and yet little deviation for 10 TeV gamma rays, since the
standard soft target threshold $m^2/E$ is half as large for the 20 TeV
gamma rays.

To formulate the constraints we begin by identifying the contour
in the $\xi$--$\eta$ plane, for which the threshold is not shifted
away from the Lorentz-invariant value. For $n=2$ this no-shift
contour is given by the diagonal $\xi=\eta$ (corresponding to
equal speeds of light for electrons and photons), which is
independent of the soft photon energy $\omega_0$. For $n=3$ the
contour is given by the joined symmetric and asymmetric $\b=1$
contours (\ref{symmn3}) and (\ref{asymmn3}) converted to the
unscaled parameters,
\begin{eqnarray}
\xi &=& \displaystyle{\frac{\eta}{2}}
\qquad\qquad\qquad\qquad\qquad\mbox{for
$\eta>-8\omega^{3}_{0}/m^4$} \label{kstsy}\\ \nonumber
\\
\xi &=& \displaystyle{\eta-\frac{4\omega_{0}^3}{m^4}+
\sqrt{-\frac{8\omega_{0}^3}{m^4}\eta}} \qquad\mbox{otherwise}
\label{kstasy}
\end{eqnarray}
The symmetric part is independent of $\o_0$ but the joining point
and the asymmetric part are not.

Above the no-shift contour, Lorentz violation {\it lowers} the
threshold. Since the shift would be larger for higher energy gamma
rays this might, depending on the details of the IR background
spectrum, enhance the ``pile-up" in the reconstructed source spectrum
if the IR backgrounds of \cite{Protheroe:2000hp} are used, or it might
produce a pile-up where one did not otherwise exist if the IR
background of~\cite{stecker01} is used.  We thus consider it unlikely
that there is much downward shift of the threshold. In any case,
nearly all of the region above the no-shift line is already excluded
by the photon decay and \v{C}erenkov constraints.

Below the no-shift contour, Lorentz violation {\it raises} the
threshold. We now consider the constraints this can yield in the cases
$n=2$ and $n=3$.

\paragraph{$n=2$ Photon annihilation constraints.}
Constraints in the $n=2$ case have been previously examined in
Ref. \cite{SG01}, although it was not realized there that the maximum
upper shift is $\b=2$, beyond which the process does not occur at
all. The $\b=2$ contour (\ref{n2contour}) is a line of unit slope and
$\widetilde{\xi}$--intercept $-1$ in the scaled parameters, hence unit
slope and $\xi$--intercept $-\o_0^2/m^2\sim -10^{-15}$. As long as the
25 meV photons annihilate at least with 20 TeV photons (whose normal
threshold is 12.5 meV), the parameters must lie above this line.

\begin{figure}[htb]
\vbox{ \vskip 8 pt
\centerline{\includegraphics[width=2.4in]{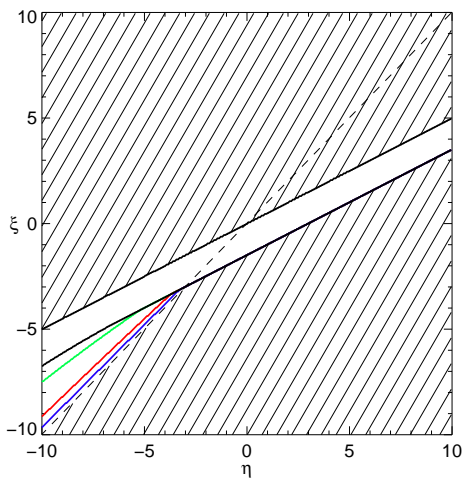}}
\caption{\label{fig:gg-n3-ph} The unfilled region indicates
parameters allowed if the lower threshold for a soft photon of 25
meV is $(a)$ not shifted down and $(b)$ not shifted up by more
than 1.5, 2, 5, 10, and infinity. The upper line is the no-shift
contour. No curvature due to the asymmetric solution is visible
for this line because the junction point as defined in
Eq.~(\ref{kstsy}) is at $\eta=-20$. The line for the existence of
a lower threshold is the lowest line. It is coincident with the
symmetric $\b=1.5$ line below the diagonal and with the (dashed)
diagonal below the crossing point. The curves stemming from the
$\b=1.5$ contour are the asymmetric contours for $\b=10,5,2$, with
lower values of $\b$ corresponding to the curves with less slope.
\smallskip}
}
\end{figure}

\paragraph{$n=3$ Photon annihilation constraints.}
For $n=3$ the contours of constant threshold in the scaled parameters
$\widetilde{\eta}$ and $\widetilde{\xi}$ are shown in
Fig.~\ref{fig:ggstruct}. The process does not occur for parameters
below a broken line consisting of the diagonal up to
$\widetilde{\eta}=\eta \times m^4/\o_0^3=-32/27$, and the line of
slope $1/2$ for greater $\widetilde{\eta}$. If absorption at $\o_0$ is
occurring for {\it any} hard gamma ray, the parameters must lie above
this broken line, so in particular everything on and below the
diagonal is excluded for $\widetilde{\eta}<-32/27$.  For $\o_0=25$ meV
this corresponds to $\eta<-2.3\cdot 32/27\approx -2.7$. This is
important, since it is a strong constraint excluding most of the
diagonal, which has been preferred by some researchers~\cite{ACP,
Aloisio:2000cm}. It is likely that a much stronger constraint holds
however, restricting the lower threshold at 25 meV to be not more than
some number of order unity times its usual value. We have indicated in
Fig.~\ref{fig:gg-n3-ph} the form of the region below the no-shift
contour and above the shift-less-than-$\b$ contour for $\b$ equal to
10, 5, 2 and 1.5. A stronger constraint would not exclude more of the
diagonal, but it has the potential to chop off the infinite wedge of
Figure~\ref{fig:cergdec} at around the same place it excludes the
diagonal.
%-------------------------------------------------------------------
%------------------------------------------------------------------------

%%%%%%%%%%%%%%%%%%%%%%%%%%%%%%%%%%%%%%%%%%%%%%%%%%%%%%%
\subsection{QED processes without thresholds}
\label{sec:qednoth}
%%%%%%%%%%%%%%%%%%%%%%%%%%%%%%%%%%%%%%%%%%%%%%%%%%%%%%%

We now consider two QED effects that occur in the presence of
Lorentz violation without any threshold, velocity dispersion of
photons in vacuo and photon splitting. The former will eventually
provide competitive constraints on $\eta$ and $\xi$ respectively,
but the latter has too slow a rate to be important.

%-------------------------------------------------
\subsubsection{Velocity dispersion of photons}
\label{sec:veldisp}
%--------------------------------------------------

Gamma-ray bursts (GRB's) are explosive extragalactic
events that release a large number of high energy photons
with a flux that varies rapidly in time. It was therefore
realized~\cite{Acea,Ellis:1999sd} that they can provide
interesting constraints or possible observations of
Planck scale suppressed Lorentz violation in the
dispersion relation for photons (a possibility noted long
ago in~\cite{Pav}).  The reason is that
while propagating over such a long distance even tiny
differences in group velocity could produce detectable
time differences between the arrival at Earth of photons
of different energy.

For photons with Lorentz breaking dispersion relations of order
$n$, $\xi$ is related to the fractional variation in group
velocity by
\begin{equation}
\xi=\frac {2} {n-1} \frac {M^{n-2}} {k_1^{n-2} -
k_2^{n-2}}\frac{\Delta c}{c}.
\end{equation}
An upper limit on the difference in arrival times of
photons from the same event provides an upper limit on
the relative speed difference, if one assumes there is no
conspiracy of different emission times cancelling
different propagation times. Together with the energies
of the different photons, such observations provide a
constraint on $|\xi|$.

The strongest constraint available today comes from GRB
930131~\footnote{
%-------------------------------------------------------
Sarkar~\cite{Sarkar:2002mg} has criticized the use of
this particular gama ray burst since this object has no
measured redshift, and hence an uncertain distance. Other
bursts~\cite{Ellis:1999sd} or blazar flares~\cite{Biller}
  give somewhat weaker constraints.
%-------------------------------------------------------
}, a gamma ray burst at a distance of 260 Mpc that
  emitted gamma rays from 50 keV to 80 MeV on a timescale
  of milliseconds~\cite{sommer}.
  Schaefer~\cite{schaefer} finds the upper limit $\Delta
  c/c<9.6\cdot 10^{-19}$ for photons of energy
  $k_1=78.6$~MeV, and $k_2=30$~keV. This yields the
  constraint $|\xi|<122$ for $n=3$.  This is weaker than
  the constraint we have from photon annihilation, hence
  time of flight data do not at present strengthen our
  constraints for $n=3$. For $n=4$ dispersion the bound
  on $|\xi|$ is on the order of $|\xi|<10^{18}$, so we
  get no interesting constraint for $n>3$.  The situation
  for $n=3$ will be significantly improved in the future
  thanks to GLAST, the gamma ray large area space
  telescope, which should be able to set limits of order
  unity on $\xi$~\cite{Norris:1999nh}.

%------------------------------------
\subsubsection{Photon Splitting}
%------------------------------------

The photon splitting processes $\gamma \rightarrow 2
\gamma$ and $\gamma \rightarrow 3 \gamma$, etc.\ do not
occur in standard QED.  Although there are corresponding
Feynman diagrams (the triangle and box diagrams), their
amplitudes vanish.  In the presence of Lorentz violation
these processes are generally allowed when
$\xi>0$. However, the effectiveness of this reaction in
providing constraints depends heavily on the decay
rate. We now give an estimate of this rate, independent
of the particular form of the Lorentz violating theory,
which indicates that the rate involves at least four
Lorentz violating factors, so is apparently too small to
be relevant at observed photon energies.

We carry out the analysis allowing for any terms in the
amplitude consistent with gauge and translation
invariance.  The particular form of Lorentz violation
considered in this paper also preserves rotation
invariance in a preferred frame, however the following
argument will not use that condition. Since gauge
invariance is preserved, the amplitude for the process
$\gamma \rightarrow N \gamma$ should arise from a term
that is a scalar formed from $N$ factors of the
electromagnetic field strength $F_{ab}$ corresponding to
the external photon legs. For each photon, $F^{({\rm
    s})}_{ab}\sim k_{[a}\epsilon_{b]}$, where $k_a$ is
the 4-momentum and $\epsilon_b$ is the polarization
vector.

In the Lorentz invariant case the equations of motion imply that
$k_a$ is a null vector and $k_a\epsilon^a=0$. Energy-momentum
conservation then implies that these 4-momenta are all parallel,
so being null they are orthogonal to each other and to all the
polarization vectors. The rate thus vanishes for two different
reasons. First, since the momenta are necessarily  all parallel,
the phase space has vanishing volume. Second, the rate must be a
scalar formed by contracting these four field strengths using only
the metric. Any such contraction vanishes since it must involve
contractions of the momenta with each other or with the
polarizations. Hence the amplitude vanishes. In the case of an odd
number of photons, another reason for  vansihing of the amplitude
is Furry's theorem, which states that the sum over loops with an
odd number of electron propagators vanishes.

If there is Lorentz violation then none of the above reasons for a
vanishing rate apply.  First of all the $N$-odd amplitudes are no
more guaranteed to vanish. Indeed for sufficiently general
implementations of Lorentz violation the Furry theorem can be
violated (see e.g.~the discussion of the Furry theorem and its
violation in the extended QED~\cite{Kostelecky:2001jc}). Secondly,
the contractions of the field strengths might involve not just the
metric but also a Lorentz violating tensor (for example $u^a u^b$
in the rotation invariant case, where $u^a$ is the unit timelike
vector specifying the preferred frame.) Finally,  in the presence
of Lorentz violation the photon four-momenta are in general not
null vectors hence they need not be parallel and they need not
vanish upon contraction. (To satisfy energy-momentum conservation
$\xi$ must be positive.)

In order for the phase space to not have vanishing volume, at
least one of the 4-momenta must involve a Lorentz-violating factor
$\d=\xi (k/M)^{n-2}$. This is not enough for the amplitude to not
vanish however. For $\g\rightarrow N\g$ with $N=3$ or 4 the
contraction of the 3 or 4 field strength tensors $F^{({\rm
s})}_{ab}\sim k_{[a}\epsilon_{b]}$ using only the metric involves
at least two vanishing contractions, and for larger $N$ there are
more. One of those vanishing contractions can be rendered nonzero
by the single Lorentz violating factor already invoked on an
external photon momentum, but the other one requires either
another such factor, or a Lorentz violating tensor in the operator
whose matrix element is being computed. Such a tensor comes with
some coefficient with dimensions determined by the dimension of
the operator. We also use the symbol $\d$ to indicate this sort of
Lorentz-violating factor.

The possible contributions to the amplitude  will therefore be
suppressed by at least two factors of $\d$. The rate goes like the
square of the amplitude, hence we infer that at energies well
above the electron mass the decay rate must behave as $E\d^4$ or
slower, where $E$ is the initial photon energy. (There is an
additional factor of $\a^N$ if we consider standard QED diagrams
for which each external photon leg comes with a factor of the
electric charge in the amplitude.)

The lifetime is therefore at least of order $\d^{-4}E^{-1}$, which
for a photon of energy 50 TeV is $10^{-29}\d^{-4}$ seconds. Such
50 TeV photons arrive from the Crab nebula, about $10^{13}$
seconds away, so the best constraint (i.e. if there is is no
further small parameter such as $\a^N$ or $1/16\pi^2$ in the decay
rate) we could possibly get on $\d$ from photon splitting is
$\d\lesssim 10^{-10}$. For $n=2$ this is not competitive with the
other constraints already obtained. For higher $n$, each
contribution arising from an operator of dimension greater than
four will be suppressed by at least one inverse power of the scale
$M$. For example, the contributions from $n=3$ deformations to the
dispersion relation will yield $\d\sim\xi E/M$. In this case the
strongest conceivable constraint on $\xi$ would be of order
$\xi\lesssim 10^4$, and even this is not competitive with the
other constraints we have found.

%----------------------------------------------------------------
\subsection{Combined Constraints}
%----------------------------------------------------------------

Having completed our discussion of photon--electron processes we
now turn to the determination of the global constraints that can
be derived from the combination of all the above results. The
photon splitting and the time of flight constraints are not as
strict as those determined by the other considered interactions,
at least for quadratic and cubic deformations, although in the
future time of flight constraints may become competitive.

%----------------------------
\subsubsection{n=2}
%----------------------------

In the case of quadratic deviations only the difference $\xi-\eta$
is constrained. The vacuum \v{C}erenkov effect yields
$\xi-\eta>-10^{-17}$, while photon decay provides the constraint
$\xi-\eta<10^{-16}$.  Together these confine $\xi-\eta$ to a small
neighborhood of zero. The photon annihilation ``likelihood
region'' would just impose $\xi-\eta\lesssim 10^{-15}$, which does
not further strengthen the constraint.

%----------------------------
\subsubsection{n=3}
%----------------------------

\begin{figure}[htb]
\vbox{ \vskip 8 pt
\centerline{\includegraphics[width=2.7in]{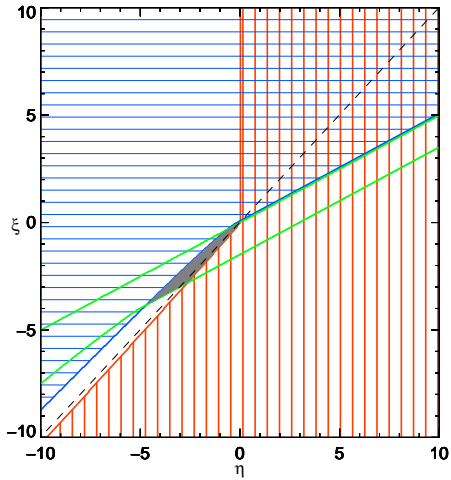}}
\caption{Combined constraints on the photon and electron
parameters, for the case $n=3$. The regions excluded by the photon
decay and \^{C}erenkov constraints are lined horizontally in blue
and vertically in red respectively.  The region between the two
diagonal green lines is where the threshold for the annihilation
of a gamma ray with a 25 meV photon ranges from its standard value
(upper diagonal green line) to not more than twice that value.
The shaded patch is the part of the allowed region that falls
between these gamma annihilation thresholds. The dashed line is
$\xi=\eta$.~\label{fig:all}
\smallskip}
}
\end{figure}

Putting together the constraints from the three photon--electron
interactions previously considered we obtain a remarkably small
allowed region in the $\eta$--$\xi$ plane (see
Figure~\ref{fig:all}).
%----------------------------------------------------------------------
%------------------------------------------------------------------------
The photon decay and \v{C}erenkov constraints exclude the
horizontally and vertically filled regions respectively. The
allowed region lies in the lower left quadrant, except for an
exceedingly small sliver near the origin with $0<\eta\lsim
10^{-3}$ and a small triangular region ($-0.16\lsim\eta<0$,
$0<\xi\lsim 0.08$) in the upper left quadrant. The discussion of
the photon annihilation threshold in subsection~\ref{sec:photann}
indicates that, although no firm constraint can be given at
present, the allowed region cannot lie too far from the corridor
between the two roughly parallel diagonal lines. These lines
indicate where the threshold for the annihilation of a gamma ray
with a 25 meV photon ranges from its standard value (upper
diagonal green line) to not more than twice that value.

If future observations of the blazar fluxes and the IR background
yield agreement with standard Lorentz invariant kinematics, the
region allowed by the photon annihilation constraint will be
squeezed toward the upper line ($k_{\rm th}\approx k_{\rm s}$).

Time of flight constraints for high energy photons currently
constrain $\xi$ to be less than $\sim 100$ at best, but future
observations should allow such constraints to further narrow the
allowed region towards the origin.

%----------------------------
\subsubsection{n=4}
%---------------------------

The case of quartic deviations is unfortunately just mildly
constrained from the available observations. The order of
magnitude allowed for the parameters is as small as $10^{11}$
(from \v{C}erenkov) for the electron--photon vertex interactions.

%----------------------------------------------------------------
\section{Interactions with protons, neutrinos, and muons}
\label{sec:other}
%----------------------------------------------------------------

We have focused so far on effects involving just electrons and
photons, in order to determine the strongest available combined
constraints. We now briefly discuss some other interactions that
are realizable with a violation of Lorentz invariance, and which
can now or in the future provide further constraints or
observations of Lorentz violation.

%--------------------------------------------------------------
\subsection{Alternative vacuum \v{C}erenkov effects:
protons, neutrinos and muons}
%-------------------------------------------------------------

The former discussion of the vacuum \v{C}erenkov effect can be
applied also for any other particle that couples to photons, using
the same kinematic equations. Since the strength of the
observational constraint is determined by the smallness of the
ratio $m^2/p^{n}_{\rm max}$, smaller masses or larger energies
generally lead to stronger constraints. However, in the case of
neutral particles that couple to photons only through higher
multipole moments the {\it rate} must also be considered.
We summarize in Table~\ref{tab:ceren} the values of the quantity
$m^2/p_{max}^n$.

%%%%%%%%%%%%%%%%%%%%%%%%%%%%%%%%%%%%%%%%%%%%%%%%%%%%%%%%%%%%%%%%%%%%%%%%%%
\begin{table*}[ht]
\caption{Typical values for different particles for the actual or
potential constraints from absence of the vacuum \v{C}erenkov
effect.\label{tab:ceren}}
%\begin{ruledtabular}
\begin{tabular}{c||l|l|l|l|l|l|l|l}
\hline\hline {} &
\multicolumn{2}{c|}{\raisebox{0pt}[13pt][7pt]{$\nu$}} &
\multicolumn{2}{c|}{\raisebox{0pt}[13pt][7pt]{$e^{-}$}} &
\multicolumn{2}{c|}{\raisebox{0pt}[13pt][7pt]{$\mu^{-}$}}&
\multicolumn{2}{c}{\raisebox{0pt}[13pt][7pt]{$p^{+}$}} \\
\hline\hline {$m$} &
\multicolumn{2}{c|}{\raisebox{0pt}[13pt][7pt]{$\lesssim 1$ eV}}&
\multicolumn{2}{c|}{\raisebox{0pt}[13pt][7pt]{$ 0.511$ MeV}}&
\multicolumn{2}{c|}{\raisebox{0pt}[13pt][7pt]{$ 105$ MeV}}&
\multicolumn{2}{c}{\raisebox{0pt}[13pt][7pt]{$ 938$ MeV}}\\ \hline
{$p_{\rm max}$} & \multicolumn{2}{c|}{\raisebox{0pt}[13pt][7pt]
{$\sim 1$ TeV -- $10^{20}$ eV~\footnote{Lower value is AMANDA
data; largest value is potentially observable UHE neutrinos.}}}&
\multicolumn{2}{c|}{\raisebox{0pt}[13pt][7pt]{$\sim 100$
TeV~\footnote{Energy expected for electrons responsible for the
creation of $\sim 50$ TeV gamma rays via inverse Compton
scattering~\cite{Koyama,Kifune:1999ex}.} }}&
\multicolumn{2}{c|}{\raisebox{0pt}[13pt][7pt]{$\sim 1$
PeV~\footnote{ Expected energies to be detected for muons produced
by cosmic neutrinos.} }}&
\multicolumn{2}{c}{\raisebox{0pt}[13pt][7pt] {$\sim 5\cdot
10^{19}$ eV~\footnote{Detected in UHECR.} }}\\ \hline {} &
\raisebox{0pt}[13pt][7pt]{$\:n=2\:$}&
\raisebox{0pt}[13pt][7pt]{$\:\:\sim 10^{-24}$ -- $10^{-40}\:\:$} &
\raisebox{0pt}[13pt][7pt]{$\:n=2\:$}&
\raisebox{0pt}[13pt][7pt]{$\:\:\sim 3 \cdot 10^{-17}\:\:$} &
\raisebox{0pt}[13pt][7pt]{$\:n=2\:$}&
\raisebox{0pt}[13pt][7pt]{$\:\:\sim 10^{-14}\:\:$} &
\raisebox{0pt}[13pt][7pt]{$\:n=2\:$}&
\raisebox{0pt}[13pt][7pt]{$\:\:\sim 4\cdot 10^{-22}\:\:$}\\
\cline{2-9} \raisebox{0pt}[13pt][7pt]{${m^{2}}/{p_{\rm max}^{n}}$}
& \raisebox{0pt}[13pt][7pt]{$\:n=3\:$}&
\raisebox{0pt}[13pt][7pt]{$\:\:\sim 10^{-8}\:\,$ --
$10^{-32}\:\:$} & \raisebox{0pt}[13pt][7pt]{$\:n=3\:$}&
\raisebox{0pt}[13pt][7pt]{$\:\:\sim 3 \cdot 10^{-3}\:\:$} &
\raisebox{0pt}[13pt][7pt]{$\:n=3\:$}&
\raisebox{0pt}[13pt][7pt]{$\:\:\sim 10^{-1}\:\:$}&
\raisebox{0pt}[13pt][7pt]{$\:n=3\:$}&
\raisebox{0pt}[13pt][7pt]{$\:\:\sim 8\cdot 10^{-14}\:\:$}\\
\cline{2-9} {}& \raisebox{0pt}[13pt][7pt]{$\:n=4\:$}&
\raisebox{0pt}[13pt][7pt]{$\:\:\sim 10^{8}\;\,\,$ --
$10^{-24}\:\:$} & \raisebox{0pt}[13pt][7pt]{$\:n=4\:$}&
\raisebox{0pt}[13pt][7pt]{$\:\:\sim 3 \cdot 10^{11}\:\:$} &
\raisebox{0pt}[13pt][7pt]{$\:n=4\:$}&
\raisebox{0pt}[13pt][7pt]{$\:\:\sim 10^{12}\:\:$} &
\raisebox{0pt}[13pt][7pt]{$\:n=4\:$}&
\raisebox{0pt}[13pt][7pt]{$\:\:\sim 2\cdot 10^{-5}\:\:$}\\
\hline\hline
\end{tabular}
%\end{ruledtabular}
\end{table*}
%%%%%%%%%%%%%%%%%%%%%%%%%%%%%%%%%%%%%%%%%%%%%%%%%%%%%%%%%%%%%%%%%%%%%%%%%%

\subsubsection{Protons}\label{sec:alt-cer-p}
Very strong constraints can be obtained using the ultra
high energy protons in cosmic rays, up to the GZK cutoff
of $5\cdot 10^{19}$ eV. The identity of these particles
has been called into question by the candidate events
beyond the GZK cutoff as described in
Sect.~\ref{sec:GZK}. However, even if the highest energy
events do not originate with protons, there is strong
evidence that protons up to the GZK cutoff do exist in
cosmic rays~\cite{Bahcall:2002wi}.\footnote{{\em Note
    added in proof}. A recent analysis~\cite{DeMarco}
  argues that there are insufficient statistics to
  establish the GZK cutoff at this time, hence the
  existence of these protons cannot yet be regarded as
  established.}

The rate of vacuum \v{C}erenkov radiation from charged particles
is irrelevant for the determination of constraints since it is
very high. (See Sect. \ref{sec:obcerenkov}.)  For the parameter
region where the threshold occurs with emission of a zero energy
photon, the proton can presumably be treated as a point charge so
the threshold relations previously obtained for electrons are
directly applicable using the proton mass in place of the electron
mass, and the parameter $\eta_p$ from the proton dispersion
relation in place of $\eta_e$. This region of parameter space is
described in section \ref{sec:cerenkov}.

For parameters where a hard photon is emitted at threshold, the
role of the partonic structure of the proton needs to be examined,
which we have not done. It may turn out that the threshold can be
determined by the quark dispersion relation rather than that of
the proton. If so, it would be the quark deformation parameter
$\eta_q$ rather than $\eta_p$ that is constrained by observations
of non-decaying high energy protons, and one would need to use the
quark mass and energy in the threshold relations. In this case the
proton may be destroyed rather than just slowed by vacuum
\v{C}erenkov radiation, however that distinction is irrelevant for
the determination of constraints, since either way high energy
protons would not travel long distances.

In estimating constraints we ignore here the possible role of
partonic structure, and simply use the proton mass and energy in
the threshold formulae derived in section \ref{sec:cerenkov} for
point particles, with the understanding that for hard emission
thresholds the constrained parameter may be $\eta_q$ rather than
$\eta_p$, and the numbers may be off by a few orders of magnitude
since the quark mass and energy were not used.

Using the GZK cutoff ($5\cdot 10^{19}$ eV) for the highest energy
protons we obtain the following constraints relating the parameter
$\xi$ in the photon dispersion relation and $\eta_{p}$ in the
proton dispersion relation.  For a quadratic deformation of the
dispersion relation ($n=2$) the bound is $\eta_p-\xi<4\cdot
10^{-22}$.  For cubic deformations ($n=3$) the constraints on
parameter space have the same form as represented in
Figure~\ref{fig:cer-n3-ph}. In the case of the proton the quantity
$m^{2}_{p}/p_{\rm max}^{3}$, is of order $10^{-14}$ compared with
$10^{-3}$ in the case of $100$ TeV electrons, which means that the
boundaries of the allowed region are closer to the $\xi$ axis in
the upper half plane and to the diagonal in the lower half plane.
However, the qualitative nature of the allowed region is
identical. A good constraint is even obtained for the case of
quartic ($n=4$) deviations. As shown in
section~\ref{sec:cerenres}, it is the quantity $m^{2}_{p}/(p_{\rm
max}^{4})$ that determines the strength of the constraint in this
case. For $5\cdot 10^{19}$ eV protons this is approximately
$10^{-5}$, still much less than unity and a much better figure
than the $10^{11}$ obtained for the $100$ TeV electron. For $n=5$
deviations the strength of the constraint is determined by
$m^{2}_{p}/(p_{\rm max}^{5})\sim 10^{3}$, hence one does not
obtain even order unity constraints on the coefficients.

%-------------------------------------%
\subsubsection{Neutrinos}
%-------------------------------------%

In the standard model the vacuum \v{C}erenkov reaction with
neutrinos, $\nu \rightarrow \nu + \gamma$, is not allowed due to
energy-momentum conservation - whether or not the neutrinos are
massive.  If they are massive energy-momentum conservation cannot
be satisfied at all. If they are massless it can only be satisfied
if all three particles  are strictly parallel, yielding no phase
space for the reaction.   (Since there is good evidence that
neutrinos have mass, we will assume this for the rest of the
discussion.)  Energy-momentum conservation is the only obstruction
for this reaction,  since although the neutrino is neutral there
is a nonzero matrix element for the process. In particular there
are two channels: the charge radius interaction and, if massive, a
magnetic moment interaction (see e.g.~\cite{Rabi}).  We therefore
see that, as for charged leptons, Lorentz violating dispersion
relations can allow the reaction to happen.

In order for the neutrino \v{C}erenkov reaction to give strong
constraints on Lorentz violation two conditions must be satisfied: (1)
the energies where Lorentz violating terms are comparable to the
neutrino mass term in the dispersion relation must be accessible to
observation, and (2) the rate of the reaction must be high enough so
that it would significantly affect the propagation of observed
neutrinos. The first condition is already met since the relevant
energy where Lorentz violation becomes important is $100$ MeV (see
Table \ref{tab:en}), while Super Kamiokande has detected neutrinos
over $100$ GeV~\cite{SuperK} and the AMANDA detector has seen
neutrinos up to a few TeV~\cite{Ahrens:2002gq}.  The second condition
is more problematic since both the charge radius and magnetic moment
channels are very strongly suppressed.

The best case for current observations would be using AMANDA, since
the neutrinos have the highest energy and travel the diameter of the
earth after being produced in the atmosphere above the North Pole. We
have not carried out a detailed analysis, but an estimate given below
suggests that the \v{C}erenkov rate is not high enough to produce
interesting constraints with these neutrinos.  The energy loss rate
depends strongly on the energy however, so atmospheric PeV neutrinos,
which are likely to be detected by AMANDA or IceCube
(seee.g.~\cite{AMANDA}), may provide constraints. (The same
experiments should detect PeV muons as secondary products of the
neutrinos, which would also provide an interesting constraint as seen
in Table \ref{tab:ceren}.)  Still higher energy neutrinos, up to
perhaps $10^{20}$ eV, are expected either as cosmic ray primaries or
as a byproduct of cosmic rays~\cite{G,Stecker79}. Such high energy
neutrinos could be detected by AMANDA~\cite{Hundertmark}, and they
could be observed via horizontal or possibly upward air showers using
existing detectors like HiRes or future ones such as the Telescope
Array~\cite{Kusenko:2001gj,Feng:2001ue}.

In Table~\ref{tab:ceren} we summarize the typical constraints one can
expect from neutrinos in the above mentioned range of energies
provided the rate is high enough. Remarkably, the combination of high
energies and low mass could give for cosmological neutrinos
($E_{\nu}\gg 1$ PeV) stringent constraints ($\eta_{\nu}\ll 1$) for
deviations up to $n=6$.

A calculation of the neutrino vacuum \v{C}erenkov rate is beyond the
scope of this article but we provide here a rough estimate that may
provide some guidance in this problem.  We saw before
(Section~\ref{sec:cerenkov}) there are two types of Cerenkov
thresholds depending on the values of $\xi$ and $\eta_\nu$: the
``soft'' one which occurs with emission of a zero energy photon, and
the ``hard'' one in which a photon with energy comparable to the
incoming particle is emitted.  The decay rate will be much greater in
the hard threshold case, so we consider that here. (The soft threshold
case may still be relevant well above threshold.)

To show that the rate might be fast enough to provide a useful
constraint it suffices to examine the charge radius interaction.  This
occurs via the emission of a virtual W-boson, hence the amplitude goes
like $\kappa k_4^2/M_W^2$, where $\kappa$ is a small numerical factor
($\sim\! 10^{-6}$) coming from coupling constants and integration
measure, $k_4^2$ is the square of the photon four-momentum, and $M_W$
is the W-boson mass. We thus estimate the rate for \v{C}erenkov
emission from a neutrino of very high energy $E$ to be $\Gamma\sim
(\kappa k_4^2/M_W^2)^2 E_{\nu}$. (The factor of $E_{\nu}$ is
determined by the phase space integration, which does not involve any
Lorentz violating factors well above threshold.) With Lorentz
violating dispersion of order $n$ we have $k_4^2=\xi k^n$, hence the
rate goes like $\Gamma\sim (\kappa \xi k^n/M_W^2)^2
E_{\nu}$.~\footnote{
%-----------------------------
Note that we cannot constrain $\xi$ as much as the threshold would
indicate, since for extremely small $\xi$ the decay rate
eventually gets too small.}
%-----------------------------
Taking the photon energy to be of the same order as the neutrino
energy $k\sim E_{\nu}$, this gives a lifetime for emission $\tau\sim
\xi^{-2} (E_{\nu}/{\rm PeV})^{-(2n+1)}\times 10^{26n-86}$ seconds.  If
correct this would be short enough to yield interesting constraints
for $n=3$ using atmospheric PeV neutrinos travelling through the
earth, since their transit time is of order $10^{-2}$ s.

As a final remark, we note that the related process of photon decay to
two neutrinos could also take place in the presence of Lorentz
violation.  This would yield strong constraints on $\xi$ and
$\eta_{\nu}$ provided the rate is high enough. The above estimate
suggests that for multi-TeV photons from cosmological sources the rate
would indeed be high enough for $n=2,3$.

%----------------------------------------------------------------%
\subsection{The GZK cutoff}
\label{sec:GZK}
%-----------------------------------------------------------------%
The presence of the GZK cutoff on the ultra high energy (UHE) proton
spectrum is due to pion photoproduction: $\gamma\, p \to p\, \pi^0$,
as previously discussed in the Introduction. The observation of this
cutoff also gives constraints on the Lorentz violating
coefficients. Current data from the HiRes, Fly's Eye and Yakutsk
experiments strongly indicate that the GZK cutoff is present at a
cosmic ray energy of $5\cdot 10^{19}$ eV~\cite{Bahcall:2002wi}. While
AGASA reports a number of extra events beyond the expected flux of
high energy cosmic rays above $10^{20}$ eV, below $10^{20}$ eV AGASA
also shows evidence for the GZK cutoff (see e.g. Figure 1
of~\cite{Bahcall:2002wi}).  Unfortunately, the experimental data are
strongly affected by the uncertain energy calibration of each
experiment.  A systematic analysis of the data allowing for various
calibrations is outside the scope of this work, so for now we assume
that the published energy calibrations are correct.

We constrain $\eta_p, \eta_\pi$ by determining where the induced
modification of the cutoff would disagree with the data.  (The
incoming photon has low energy and so no useful constraints on $\xi$
are obtained.) The $\eta_p, \eta_\pi$ constraints are quite strong (on
the order of $10^{-10}$ for $n=3$) due to the high energy of the
reaction.

In the standard Lorentz invariant theory the threshold energy for pion
production is $E_{\rm th}= m_{\pi}(2\,m_{p}+m_{\pi})/4\o_{0}$, so a
photon with energy $\omega_{0} \sim 1.3$ meV is at threshold with the
proton at the GZK energy. In order to give a constraint on Lorentz
violations we consider raising or lowering the UHE proton at threshold
with the same $\omega_{0}$. This is equivalent to changing the GZK
cut-off as we are modifying the UHE proton energy that interacts with
the relevant CMBR photons responsible for the Lorentz invariant GZK
effect.

Examination of the data plot in Fig. 2 of Ref.~\cite{Bahcall:2002wi}
reveals that if the cutoff were shifted via a Lorentz violating effect
down to $2\cdot 10^{19}$ eV or up to $7\cdot 10^{19}$ eV then the
theoretical predictions would no longer agree with the data at above a
$2 \sigma$ confidence level.  This energy range therefore provides
constraints on $\eta_p, \eta_\pi$.  From the threshold theorems of
\cite{JLMth} we again know that in the threshold configuration where
the GZK reaction begins to occur the incoming proton and photon
collide head on and the outgoing proton and pion 3-momenta are
parallel.  Energy-momentum conservation in this configuration and the
dispersion relations give an equation similar to equation
(\ref{eq:ggscat}) for photon annihilation,
\begin{equation}
0=F(p,x):=-\frac{m^{2}_{p}}{p^{n}}(1-x)^2-\frac{m_{\pi}^{2}}{p^{n}}x+
\left( \eta_{p}+\frac{4\omega_{0}}{p^{(n-1)}} \right) x
\left(1-x\right)-\eta_p x \left(1-x\right) \left[
x^{(n-1)}+\frac{\eta_{\pi}}{\eta_{p}} \left(1-x\right)^{(n-1)}\right],
\label{eq:gzk} \end{equation}
where $x=q/p$, and $p$ and $q$ are the initial and final proton
3-momenta.

For $n=2$ the threshold analysis has already been done by Coleman and
Glashow~\cite{CG} leading to a constraint $\eta_{\pi}-\eta_{p}<5\times
10^{-24}\,[\omega/\bar{\omega}]^{2}$ for a target photon $\o$, where
$\bar{\o}=kT_{\rm CMB}=0.235$ meV.

\begin{figure}[htb] \vbox{ \vskip 8 pt
 \centerline{\includegraphics[width=2.6in]{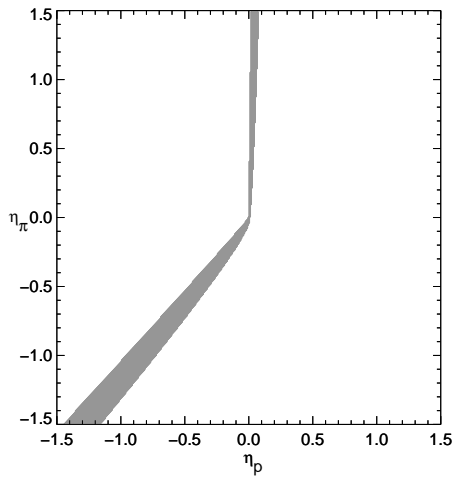}}
 \caption{\label{fig:gzk1} The range of $\eta_p, \eta_\pi$ for $n=3$
 dispersion modifications where the GZK cutoff is between $2\cdot
 10^{19}$ eV and $7\cdot 10^{19}$ eV. $\eta_p$ and $\eta_\pi$ are in
 multiples of $10^{-10}$.}
\smallskip}
\end{figure}

For $n=3$, the presence of the pion in this equation complicates the
analysis as there is an additional mass term and the final particles
are not interchangeable.  The case of equal coefficients
($\eta_{p}=\eta_{\pi}$) has been studied analytically in
~\cite{Mestres,Bertolami,ACP,Kifune:1999ex,Aloisio:2000cm} and
numerically in~\cite{Major}. Here we numerically find the thresholds
for the GZK reaction allowing for unequal coefficients in the case
$n=3$.  The region in the $\eta_p, \eta_\pi$ plane where the
thresholds are in the allowed range discussed above are shown in
Figure~\ref{fig:gzk1}, in which the axes are in multiples of
$10^{-10}$.
%--------------------------------------------------------------------
%--------------------------------------------------------------------

We turn now to the question of the extra AGASA events above $10^{20}$
eV.  The AGASA data is sparse in this energy range, and there is not a
large, precise data set from other experiments with which AGASA
disagrees. The uncertainties in all experiments are large enough that
a modified theoretical theoretical spectrum could possibly agree with
all experiments at the $1\sigma$ level.  One cannot therefore simply
disregard the possibility that the flux above $10^{20}$ eV is in fact
higher than the standard theoretical prediction.

\begin{figure}[htb]
\vbox{ \vskip 8 pt
\centerline{\includegraphics[width=2.6in]{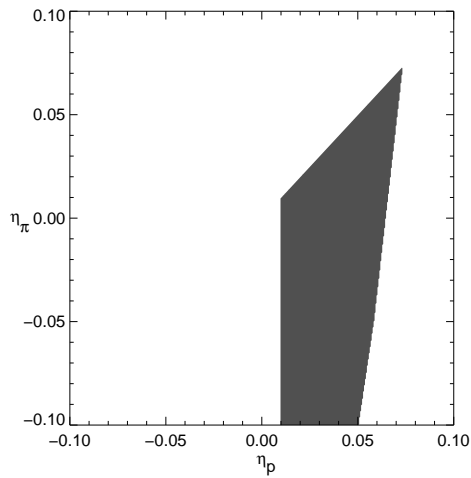}}
\caption{\label{fig:gzk2} The range of $\eta_p, \eta_\pi$ for $n=3$
dispersion modifications where the GZK cutoff is between $2\cdot
10^{19}$ eV and $7\cdot 10^{19}$ eV and the upper threshold exists
below $10^{21}$ eV. $\eta_p, \eta_\pi$ are in multiples of
$10^{-10}$. Note that this region does not include the origin, as
there is no upper threshold in the Lorentz invariant case.}
\smallskip}
\end{figure}

Previous authors have suggested that the AGASA events above the GZK
cutoff could be explained by an upward shift of the GZK cutoff induced
by Lorentz
violation~\cite{Mestres,CG,Bertolami,ACP,Kifune:1999ex,Major}, however
this seems incompatible with current data since the cutoff is
seen. Another, more subtle possibility is that these events are
related to the existence of an {\it upper} threshold. We have checked
numerically that no upper threshold exists below $10^{20}$ eV within
the allowed region of Fig.~\ref{fig:gzk1}.  Nevertheless, the phase
space for a reaction begins to close up before the upper threshold is
reached. The reduction in phase space would in turn reduce the rate of
the GZK reaction leading to a higher than expected count of events at
high energies.  If the lower threshold were dramatically modified
whenever there is an upper threshold then this scenario could not
explain the data.  However, this is not the case - there are choices
of $\eta_p, \eta_\pi$ such that an upper threshold exists and the
lower threshold is only slightly modified.  Since the lower threshold
modifications can be small, the experimental signature of the GZK
cutoff could remain unchanged near $5\cdot 10^{19}$ eV while the
intensity of the spectrum at high energies increased from its Lorentz
invariant prediction. This scenario could perhaps explain the AGASA
data and be compatible with other experiments if the upper threshold
is low enough that there is a significant phase space reduction just
above $10^{20}$ eV. The range of $\eta_p, \eta_\pi$ for which this
effect could occur is given in Figure~\ref{fig:gzk2}.
%---------------------------------------------------------------------
%--------------------------------------------------------------------
% 
There we consider the range of parameters for which the lower
threshold still lies between $2\cdot 10^{19}$ eV and $7\cdot 10^{19}$
eV but an upper threshold exists below $10^{21}$ eV (so that the
induced reduction in phase space could affect the AGASA data).

We have not considered the constraints that can be obtained from the
GZK cutoff in the case $n=4$ but it is clear that they are interesting
in this case as well, since $1/p_{\rm max} = M/p_{\rm max} = 2\cdot
10^{8}$.  The $n=3$ constraints are of order $10^{-10}$, hence the
$n=4$ ones are of order $10^{-2}$.

%---------------------------------------------------%
\subsection{Neutron stability--Proton instability}
%---------------------------------------------------%

If there are different dispersion relations for protons, neutrons,
positrons, and neutrinos then protons may be unstable and decay to
neutrons at sufficiently high energies. For $n=2$, Coleman and
Glashow~\cite{CG} have shown this explicitly. If $n>2$ then the
analysis becomes more complicated, but there exist parameters for
which the neutron is stable and the proton decays. For example,
consider a neutron with $\eta_n=-1$ and $n=3$ dispersion relation, and
an unmodified proton, electron, and neutrino. At momenta
$p_n>m_n^{2/3}$ the neutron energy-momentum vector becomes
spacelike. Since the energy momentum vectors of the other particles
are still timelike it is impossible to satisfy energy momentum
conservation hence neutron decay does not occur above this
energy. This opens the possibility that ultra high energy cosmic rays
are neutrons rather than protons, in which case the GZK cutoff for
these cosmic rays is irrelevant since neutrons do not interact
strongly with CMBR photons. The presence of the observed cutoff can
thus be used to constrain the parameters further, and one could also
contemplate the possibility that the AGASA events above the GZK cutoff
are present because the neutron becomes stable just above the cutoff.

%----------------------------------------------------------------
\section{Combined constraints for a universal Lorentz breaking dispersion
relation} \label{sec:univ}
%----------------------------------------------------------------

Although we have considered so far the case of different Lorentz
violating parameters $\eta_{a}$ for different particles, it may be
that the underlying quantum gravity physics selects a universal
deformation parameter $\eta$. This indeed was the ansatz considered in
most of the previous literature.  We therefore consider now this
special case with $n=3$ deformations.

We start by considering the photon--electron interactions.  From
Figure~\ref{fig:cer-n3-ph} and equation (\ref{eq:cerconda}) we see
that the \v{C}erenkov effect limits the available values on the
diagonal to a semi-infinite line with $\eta<m^2/2p^3_{\rm max}$. This
corresponds numerically to $\eta\lesssim 10^{-14}$ if we consider the
constraint provided by the observation of ultra-high energy protons in
cosmic rays.  The analysis of photon decay shows that the permitted
values on the diagonal $\xi=\eta$ are restricted to the semi-infinite
line $\eta<8m^2/k_{\rm max}^3\sim 8\cdot 10^{-2}$ (using the
observation of 50 TeV gamma rays from the Crab nebula).  The
observation of photon annihilation provides, as previously discussed,
a more uncertain constraint. Nevertheless for our purposes it is
enough to take into account that some absorption is detected for gamma
rays at least up to 10 TeV.  We can then take as a definite constraint
the line for the existence of a lower threshold as shown in
Figure~\ref{fig:ggstruct}. This line meets the diagonal at
$\widetilde{\eta}=-32/27$. The problem is now to decide for which
$\omega_0$ we are sufficiently confident the photon annihilation still
takes place. As a reference value we take here again the $\omega_0=25$
meV photon previously considered. In this case the region of existence
of a lower threshold for the photon annihilation limits the value of
$\eta$ to the semi-infinite range $\eta>-2.3\times(32/27)\approx
-2.7$.

If the GZK cutoff is confirmed that would establish with certainty
that at least some of the UHE cosmic rays are indeed protons.
Moreover it would also provide a correspondingly strong constraint
on negative values of $\eta$. If the GZK cutoff is within order
unity of it's Lorentz invariant value, $\eta$ is constrained to
be $|\eta| \lesssim 10^{-14}$.  (Note that this constraint on
$\eta$ is so strong as to exclude the region of upper threshold
for the GZK process shown in Figure~\ref{fig:gzk2}.)
The upper bound might be further pushed toward zero if one
takes into account the \v{C}erenkov effect of high energy neutrinos (see
Table~\ref{tab:ceren}).
%----------------------------------------------------------------

%----------------------------------------------------------------
\section{Discussion}
\label{sec:disc}
%----------------------------------------------------------------

In this paper we have performed a systematic analysis of the effects
of Lorentz violating dispersion on particle reactions, allowing for
unequal deformation parameters for different particles.  We have
analyzed the threshold kinematics and combined the observational
constraints where possible.  Even when suppressed by the inverse
Planck mass, such Lorentz violation can lead to radically new behavior
in the kinematics of particle interactions at much lower
energies. Reactions previously forbidden can be allowed, lower
thresholds can be shifted and upper thresholds can be introduced.  The
presence of upper thresholds is a feature of Lorentz breaking physics
that is not present in Lorentz invariant physics and which can be
relevant for observational constraints.\footnote{
%-------------------------------------------------------
More complex dispersion relations can lead to multiple
thresholds~\cite{JLMth} with could have further observational
effects.}
%-------------------------------------------------------
Furthermore, we have found that for interactions with identical final
particles, the final momenta can be distributed asymmetrically at
threshold. While this is a straightforward consequence of the
kinematics, it has been previously overlooked in the literature,
probably because it is alien to Lorentz invariant physics.

Using these kinematical results, we have seen that a conservative
interpretation of observations puts strong constraints on the
coefficients $\eta$ and $\xi$ of order $E/M_{\rm P}$ modifications to
the electron and photon dispersion relations. The allowed region
includes $\xi=\eta=-1$, which has been a focus of previous
work~\cite{ACP,Kifune:1999ex,Kluzniak}. The negative quadrant has most
of the allowed parameter range.  Note that in this quadrant all group
velocities are less than the low energy speed of light.  For
modifications of order $(E/M_{\rm P})^2$ there are no significant
constraints in the electron-photon sector derivable from current
observations, due to the fact that the energies of observed particles
are too low.  However reactions such as proton \v{C}erenkov (in vacuo)
or pion production by cosmic rays, for which we have data at much
higher energies, can provide good constraints for $(E/M_{\rm P})^2$
modifications (although for different particle deformation
parameters).  Ultra high energy cosmological neutrinos may also
provide good \v{C}erenkov constraints at this or even higher orders,
since the neutrino mass is much smaller than that of any other
particle. The interaction amplitudes are very suppressed however, so
it is necessary to accurately calculate the rate and compare it with
the travel time of the neutrino.

There are a number of ways to improve the constraints on $O(E/M_{\rm
P})$ modifications from electron-photon interactions.  Higher energy
electrons would not help much since the \v{C}erenkov constraint is
already strong, while finding higher energy undecayed photons would
squeeze the allowed region onto the line $\xi=\eta$ of
Figure~\ref{fig:all}. To further shrink the allowed segment of this
line would require improved knowledge of the infrared background and a
reconstruction of the source spectrum from the observed gamma rays in
the presence of Lorentz violation.  Also, the constraint from time of
flight measurement may become competitive using improved detectors.

Other constraints may be provided by additional interactions not
considered here. For example, a possible upper threshold for $e^+\,
e^- \rightarrow 2\g$ cannot provide a competitive constraint in
astrophysical observations since there are other processes by which
observed high energy photons can be produced.  However, if future
electron accelerators can reach energies above $10$ Tev then one can
expect to get a good constraint from this reaction.  In addition there
may be other reactions for which upper thresholds can produce useful
constraints at or near currently observed energies. Reactions
involving more than two types of particles, such as $\nu \rightarrow
e^- \, W^+$, could also give constraints. It may be possible that by
considering a number of such reactions a multi-dimensional parameter
space can be usefully constrained.

The idea motivating our work is that Lorentz violation may be a
consequence of quantum gravity, in which case the natural scale for
the Lorentz violation is the Planck scale. If, as in braneworld
scenarios, the quantum gravity scale were to be around a TeV, then the
natural scale for Lorentz violation induced by quantum gravity would
be the TeV scale. Clearly, the only way such Lorentz violation could
be compatible with observations is if it were extremely suppressed
compared with this natural scale. This suggests that either TeV scale
quantum gravity is wrong, or it does not violate 4d-Lorentz
invariance.

In conclusion, the {\it absence} of anomalous observations
provides stringent constraints on the possibility of Lorentz
violation originating at the Planck scale.  This in turn gives
important information as to the viability of quantum gravity
theories that predict 4-d Lorentz violation.  We can expect that,
as better data at higher energies becomes available, even stronger
constraints will be imposed or, alternatively, positive signatures
of Lorentz violation may be found. Either way, it is clear that a
useful tool for the phenomenological investigation of quantum
gravity is now at hand.

%----------------------------------------------------------------
\section*{Acknowledgments}
%----------------------------------------------------------------

We are grateful to G.~Amelino-Camelia, A.~Celotti, R.~Mohapatra,
I.Z.~Rothstein, and F.~Stecker for useful discussions.  This research
was supported in part by the National Science Foundation under Grant
No. 9800967 at the University of Maryland, Grant No. PHY99-07949 at
the KITP, and the Erwin Schr\"odinger Institute.

%-------------------------------------------------------------------
\appendix
%-------------------------------------------------------------------
\section{Photon annihilation thresholds}~\label{app:ggscat}
%-------------------------------------------------------------------

In this appendix we work out the lower and upper thresholds for
the process $\gamma \, \gamma \rightarrow e^{+} \, e^{-}$ as a
function of the Lorentz-violating parameters $\eta$ and $\xi$. Our
starting point is the kinematic equation (\ref{eq:ggscat}) derived
in section \ref {sec:gg}:
\begin{equation}
 0=F(k,y):=-\frac{m^{2}}{k^{n}}+\left( \xi+\frac{4\omega_{0}}{k^{(n-1)}} \right)
   y \left(1-y\right)-\eta y \left(1-y\right)
   \left[ y^{(n-1)}+\left(1-y\right)^{(n-1)}\right].
    \label{eq:ggscatapp}
\end{equation}
Here $m$ is the electron mass, $k$ is the magnitude of the incoming
hard photon momentum, $\omega_0$ is the soft photon energy, $y=p/k$,
where $p$ is the magnitude of the electron (or positron) momentum, and
the threshold configuration of antiparallel incoming photons and
parallel outgoing electron-positron pair has been imposed. This
equation follows from (i) energy-momentum conservation, (ii) the
dispersion relations for the particles, and (iii) the threshold
configuration. To find the lower and upper threshold for given values
of $\eta$ and $\xi$ we must determine the minimal or maximal $k$ for
which the reaction can occur. According to the threshold theorem ({\it
cf.} Sect. \ref{kinematics}) these $k$ values always occur with what
we just called the threshold configuration, hence we must determine
the minimal or maximal $k$ for which there is a solution $(k,y)$ to
the kinematic equation (\ref{eq:ggscatapp}) with $y$ in the range
$[0,1]$. In the Lorentz invariant case the threshold always occurs
with the symmetric configuration $y=1/2$, however in the
Lorentz-violating case this is not always true.

In order to derive results applicable to any value of the soft photon
energy $\omega_0$ and ``electron'' mass $m$, we introduce scaled
variables
\begin{equation}
\b=k/k_{\rm LI},\qquad \widetilde{\eta}= \eta
(m^{2(n-1)}/\omega_0^n),\qquad \widetilde{\xi}=\xi
(m^{2(n-1)}/\omega_0^n) \label{eq:scaled2}
\end{equation}
where $k_{\rm LI}$ is the standard lower threshold $m^2/\omega_0$.  In
terms of these scaled variables the equation $m^{-2}k^nF(k,y)=0$ takes
the form
\begin{equation}
    G(\beta,y)=\alpha_{n}(y)\, \b^{n}+\g(y)\, \b-1=0,
    \label{eq:gfggsapp}
\end{equation}
where
\begin{equation}
    \alpha_{n}(y)= y(1-y)\left(\widetilde{\xi}-\widetilde{\eta}\; \left[y^{n-1}+(1-y)^{n-1}\right]\right)
    \label{eq:alphaapp}
\end{equation}
and
\begin{equation}
    \g(y)=4\; y(1-y).
    \label{eq:gammaapp}
\end{equation}
Equation (\ref{eq:gfggsapp}) is a generalization of those derived by
Aloisio \etal.~\cite{Aloisio:2000cm} for the specific cases of $n=3,4$
with purely symmetric configurations ($y=1/2$) and equal deformation
coefficients ($\widetilde{\xi}=\widetilde{\eta}$).

\begin{figure}[htb]
\vbox{ \vskip 8 pt
\centerline{\includegraphics[width=2.7in]{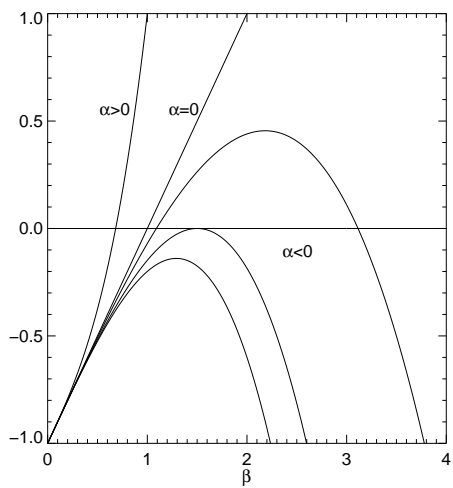}}
\caption{\label{fig:gofbeta} Behavior of $G(\beta,y)$
(\ref{eq:gfggsapp}) for fixed $y$ and different values of alpha.
\smallskip}
} 
\end{figure}

Figure \ref{fig:gofbeta} shows the general behavior of $G(\beta,y)$
(\ref{eq:gfggsapp}) for any fixed $y$ (and therefore fixed
$\gamma(y)$) and different values of $\alpha$.
%
%------------------------------------------------------------------
%------------------------------------------------------------------
%
~From the plot we see that for any $y$ there are either one, two, or
zero solutions to the kinematic equation for $\b$. If there are two
solutions, of course only the lower one is a candidate for the lower
threshold. The upper one can correspond to an {\em upper threshold},
that is, to the highest available value of $\b$ for which the reaction
is kinematically allowed. To our knowledge, the possibility of upper
thresholds has been overlooked in all of the previous literature
except for Klu\'zniak~\cite{Kluzniak} (see discussion at end of
section~\ref{subsec:uppersummary}).

For each $y$ there is a maximal value of the lower root $\b$ which
occurs when $\a_n$ is such that the curve described by
Eq.~(\ref{eq:gfggsapp}) is tangent to the $\b$ axis. This occurs
for
\begin{equation}
    \alpha^{\rm tang}_{n}=-\g^n\frac{ (n-1)^{(n-1)} } { n^{n} },
    \label{eq:alphatang}
\end{equation}
and, at this tangency point,
\begin{equation}
 \b_{\rm tang}=\g^{-1}\frac{n}{n-1}. \label{eq:btang}
\end{equation}
If $\b$ is to be a lower threshold it must lie below this tangency
point.

Given values for $\widetilde{\eta}$ and $\widetilde{\xi}$, Eq.
\ref{eq:gfggsapp} implicitly defines zero, one, or two (real positive)
solutions for $\b$ as a function of $y$. A lower threshold corresponds
to the global minimum of $\b$. Our strategy is to first find the local
minima and, if there is more than one, determine which is the global
minimum.

A local minimum of $\beta(y)$ is characterized by $\beta_y=0$ and
$\beta_{yy}>0$, where the subscript $y$ denotes derivative with
respect to $y$. To find the corresponding conditions on $G$ we use the
fact that $G(\b(y),y)$ is identically zero, hence its total derivative
with respect to $y$ vanishes: $G_\beta \beta_y + G_y = 0$, where the
subscripts on $G$ denote partial derivatives. The second total
derivative of $G(\b(y),y)$ with respect to $y$ also vanishes. At a
stationary point where $\beta_y=0$, this yields $G_\beta \beta_{yy} +
G_{yy}=0$. Thus the conditions for $\beta(y)$ to be local minimum are
\begin{equation}
\b_y=-G_y/G_\b=0\qquad\qquad {\rm and} \qquad\qquad
\b_{yy}=-G_{yy}/G_\beta>0. \label{eq:d2bdy2}
\end{equation}
It is clear from Fig. \ref{fig:gofbeta} that $G_\b$ is always
positive at the lower root of $G=0$, and vanishes only at the
tangency value of $\alpha$. Since only the lower root can be a
lower threshold, we have two necessary conditions to be a lower
threshold:
\begin{equation}
G_\b>0 \qquad\qquad {\rm and} \qquad\qquad G_{yy}<0.
\label{eq:lowerconds}
\end{equation}
We now use these considerations to find the thresholds for
$n=2,3$.

%----------------------------------------------------
\subsection{Photon annihilation thresholds for $n=2$}
%----------------------------------------------------

For $n=2$ Eq. (\ref{eq:gfggsapp}) reduces to
\begin{equation}
    G= y (1-y) \left[(\widetilde{\xi} - \widetilde{\eta}) \b^2 + 4
    \b\right]
    -1=0 \label{eq:gnequal2}
\end{equation}
There is only one extremum, at $y=1/2$. Substituting $y=1/2$ into
Eq. (\ref{eq:gnequal2}) yields
\begin{equation}
\widetilde{\xi}=\widetilde{\eta} + 4 \frac {1-\b} {\b^2}.
\label{eq:symcon2}
\end{equation}
The $\widetilde{\xi}$-intercept, $\widetilde{\xi}_0 =
4(1-\b)/\b^2$, decreases monotonically for $\b<\b_{\rm tang}$ and
increases monotonically for $\b>\b_{\rm tang}$:
$d\widetilde{\xi}_0/d\b=(\b-2)(4/\b^3)$.

The values of $\b$ less than $\b_{\rm tang}=2$ are candidates for a
lower threshold.  The contours (\ref{eq:symcon2}) for $\b<2$ are
parallel straight lines of unit slope, whose
$\widetilde{\xi}$-intercept goes monotonically from $\infty$ to $-1$
as $\b$ goes from 0 to 2. Since these lines do not cross there is only
one candidate threshold for each pair
$(\widetilde{\eta},\widetilde{\xi})$, hence these lines indeed give
the contours of the lower threshold. The reaction is forbidden below
the line $\widetilde{\xi}=\widetilde{\eta} -1$.  The highest lower
threshold is given by $k = 2k_{\rm LI}= 2m^2/\omega_0$.

The values of $\b$ greater than $\b_{\rm tang}=2$ are candidates for
upper thresholds since they are local maxima of $\b$ among the
threshold configurations. The contours for these are also given by the
straight lines (\ref{eq:symcon2}), with $\widetilde{\xi}$-intercept
that goes monotonically from $-1$ to $0$ as $\b$ goes from 2 to
$\infty$. No other candidate for the upper threshold exists at a given
value of $\widetilde{\xi}$, $\widetilde{\eta}$, so to check whether
these contours actually represent upper thresholds we need only verify
that an upper threshold exists at all.  That is, we must rule out the
possibility that there are configurations for which the annihilation
process occurs with arbitrarily large incoming photon energy. To do
this we examine the limit of large $k$.  Then the soft photon is
irrelevant, and the question is the same as whether an arbitrarily
high energy photon can decay to an electron-positron pair. We already
determined in the section on photon decay that it occurs only above
the diagonal, i.e. for $\widetilde{\xi}>\widetilde{\eta}$. Hence the
process does not occur at arbirarily large energy below the diagonal,
which is where all the candidate upper thresholds lie. Thus these
candidates are indeed all upper thresholds.

\subsection{Photon annihilation thresholds for $n=3$}

Since the two final particles are interchangeable (\ref{eq:gfggsapp})
is symmetric about $y=1/2$. For higher $n$ we can reduce the order of
this equation by introducing the variable $z=(2y-1)^2$ that also has
this symmetry, as we did for the case of photon decay in section
\ref{sec:gdecay}. The physically relevant range of $z$ is $0\le z\le
1$. With this change of variables for $n=3$ (\ref{eq:gfggsapp}) can be
written as
\begin{equation}
    G= \frac{\b^3}{4}\left[(\widetilde{\xi}+4/\b^2)(1-z)
     - ({\widetilde{\eta}}/2) (1-z^2)\right]
    -1=0.
\label{eq:gnequal3}
\end{equation}
There are now two extrema, one at $z=0$ which corresponds to the
usual symmetric $y=1/2$ case, and the other where $G_z=0$, i.e. at
\begin{equation}
z_a=\frac {\widetilde{\xi} + 4/{\b^2}} {\widetilde{ \eta}},
\label{eq:asymz}
\end{equation}
which corresponds to an asymmetric configuration in which the
outgoing particles have different momenta.

The solution of (\ref{eq:gnequal3}) for $\widetilde{\xi}$ in the
symmetric case ($z=0$) with $\b=\b_s$ yields
\begin{equation}
\widetilde{\xi}=\frac{\widetilde{\eta}} {2} + \frac {4 (1-\b_s)}
{\b_s^3}. \label{eq:bsnequal3}
\end{equation}
The $\widetilde{\xi}$-intercept, $\widetilde{\xi}_0 = 4(1-\b)/\b^3$,
decreases monotonically for $\b<\b_{\rm tang}=1.5$ and increases
monotonically for $\b>\b_{\rm tang}$
$d\widetilde{\xi}_0/d\b=(\b-1.5)(8/\b^4)$.  Only values of $\b$ less
than the tangency value 1.5 are candidates for a symmetric lower
threshold, while values greater than 1.5 are candidates for a
symmetric upper threshold.

The symmetric case can only be a lower or upper threshold when the
inequality $\widetilde{\xi}>-4/\b^2$ holds.  We can see this by
imposing the conditions for a local minimum or local maximum
respectively: $\b_{yy}>0$ when $\b<1.5$ or $\b_{yy}<0$ when
$\b>1.5$. Since $G_\b>0$ in the first case and $G_\b<0$ in the second
case (as can be seen from Fig.~\ref{fig:gofbeta}),
(\ref{eq:lowerconds}) shows that both cases require $G_{yy}<0$. To
evaluate $G_{yy}$ we note that
\begin{equation}
 d^2/dy^2 = 16z(d^2/dz^2) + 8(d/dz), \label{yyzz}
\end{equation}
hence $G_{yy}^{(s)}= 8 G_z^{(s)}= -8\b(1+ \widetilde{\xi}\b^2/4)$,
which is negative only if $\widetilde{\xi}>-4/\b^2$.

The solution of (\ref{eq:gnequal3}) in the asymmetric case
($z=z_a$) with $\b=\b_a$ yields
\begin{equation}
\widetilde{\xi}=\widetilde{\eta} - \frac {4} {\b_a^2} +
\sqrt{-\frac{8\widetilde{\eta}}{\b_a^3}} \label{eq:banequal3}
\end{equation}
The asymmetric case only exists when $\widetilde{\eta}<0$, and only
the positive square root is physically relevant, since
(\ref{eq:asymz}) gives $z_a= 1-\sqrt{-8/\widetilde{\eta}\b^3}$ and $z$
must be less than unity. Also, $z_a$ must be positive, so the
asymmetric case is only relevant when $\widetilde{\xi}<-4/\b^2$.

The asymmetric stationary point has $G_{yy}^{(a)}= 16 z_a
G_{zz}^{(a)}= 4 z_a \widetilde{\eta}\b^3<0$, hence it is a local
minimum if and only if $G_\b>0$. This corresponds to the inequality
$\widetilde{\xi}> \widetilde{\eta}+4/3\b^2$. It is a local maximum
when the opposite inequality holds. For $\b<1.5$ the asymmetric curve
represents a local minimum everywhere since it is above
$\widetilde{\xi}= \widetilde{\eta}+4/3\b^2$ everywhere in the physical
region $\widetilde{\xi}<-4/\b^2$.  For $\b>1.5$ the asymmetric curve
crosses below $\widetilde{\xi}= \widetilde{\eta}+4/3\b^2$ while still
in the physical region, where it represents a local maximum.

\begin{figure}[htb]
\vbox{ \vskip 8 pt
\centerline{\includegraphics[width=2.7in]{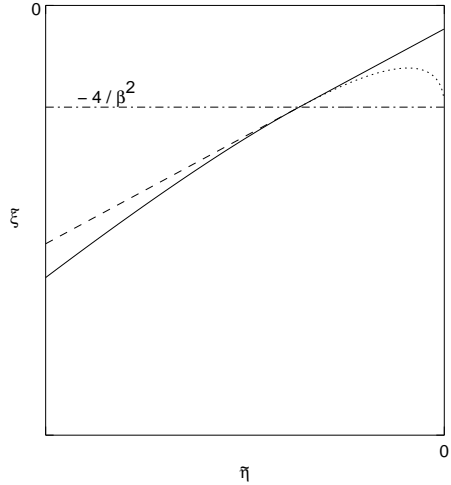}}
\caption{\label{fig:transition} The (straight) symmetric
(\ref{eq:bsnequal3}) and (curved) asymmetric (\ref{eq:banequal3})
contours for a fixed value of $\beta$. The unphysical part of the
asymmetric curve is dotted, and the part of the symmetric line that is
not a local maximum or minimum of $\b$ is dashed. The ``$\b$-curve" is
the two joined solid segments of these contours, and indcates points
where $\b$ is a candidate for the lower or upper threshold. The
joining point is
$(\widetilde{\eta},\widetilde{\xi})=(-8/\b^3,-4/\b^2)$.
\smallskip}
}
\end{figure}

The symmetric line (\ref{eq:bsnequal3}) and asymmetric curve
(\ref{eq:banequal3}) meet at
$(\widetilde{\eta},\widetilde{\xi})=(-8/\b^3,-4/\b^2)$ and are tangent
there, as shown in Figure \ref{fig:transition}.
%
%-----------------------------------------------------------------
%
%-----------------------------------------------------------------
%
Above this meeting point only the symmetric solution is a candidate
threshold, and below this point only the asymmetric solution is. As
$\b$ varies, the curve traced out by these meeting points is given by
\begin{equation}
\widetilde{\xi}_{\rm join}=- (-\widetilde{\eta})^{2/3}.
\label{eq:xijoin}
\end{equation}
We shall use the name ``$\b$-curve" for the joined curve that is
the symmetric line above and the asymmetric curve below
$\widetilde{\xi}=-4/\b^2$.

\subsubsection{Lower threshold for photon annihilation, $n=3$}

To find the contours of constant lower threshold in the
$\widetilde{\eta}$--$\widetilde{\xi}$ plane we proceed as follows.
First we choose a value of $\b<1.5$, and consider the corresponding
$\b$-curve. The points on this curve are the only candidates for the
threshold to be $\b$. To determine if the threshold actually is $\b$
at a given point we must determine whether or not there is a solution
to (\ref{eq:gnequal3}) with a {\it smaller} value $\b_<$ at the same
point. In other words, we must determine if a $\b_<$-curve could cross
the original $\b$-curve. In fact it cannot. The $\b_<$-curve starts
out above the $\b$-curve at $\widetilde{\eta}=0$ (since $(1-
\b_<)/\b_<^3 > (1- \b)/\b^3$). In the symmetric sections the slopes
are both equal to $1/2$. In the asymmetric section the slope computed
from (\ref{eq:banequal3}) is
$d\widetilde{\xi}/d\b=1-4(-8\widetilde{\eta}\b^3)^{-1/2}$, which is
always greater than $1/2$ in the region below $\xi=-4/\b^2$ and is
greater for larger $\b$ at fixed $\widetilde{\eta}$. Hence the
$\b_<$-curve is everywhere above the $\b$-curve, so the curves never
cross. The $\b$-curves with $\b<1.5$ thus give the lower threshold.

Now assume that $\b$ is greater than $1.5$, so that only the the
asymmetric configuration can be a lower threshold. In this case there
are $\b_<$-curves with smaller values of $\b_<$ that cross the
$\b$-curve. In particular, the asymmetric part of the $\b$-curve
crosses the symmetric $\b=1.5$ line from below and then goes on to
cross lines of yet smaller $\b$ above that before leaving the local
minimum region. Thus $\b$ cannot be the global minimum above the
symmetric $\b=1.5$ line, even though $\b$ remains a local minimum up
to when it crosses below the line $\widetilde{\xi}=
\widetilde{\eta}+4/3\b^2$.

The only remaining question is whether the different asymmetric
$\b$-curves for $\b>1.5$ can cross below the symmetric $\b=1.5$
line. In fact they cannot. It can be shown that the terminus on the
$\b=1.5$ line moves to larger values of $\widetilde{\eta}$ as $\b$
goes from $1.5$ to $\infty$. (See Figure~\ref{fig:ggstruct}.)  Since
the slope of the asymmetric $\b$-curves increases with $\b$ (as
discussed two paragraphs above) in the region below the symmetric
$\b=1.5$ line, the curves for different $\b$ do not cross.

In summary, we have shown that the regions where the symmetric and
asymmetric lower thresholds exist take the form shown in
Figure~\ref{fig:thstruct}. The contour lines of constant threshold are
the straight lines (\ref{eq:banequal3}) of slope 1/2 in the symmetric
region, and are given by (\ref{eq:banequal3}) in the asymmetric
region. These contours are shown in Figure~\ref{fig:ggstruct}.

%-------------------------------------------------------------

\subsubsection{Upper threshold for photon annihilation, $n=3$}

We now turn to analysis of the upper thresholds.  Our first step is to
ascertain in which region of the $\widetilde{\eta}$--$\widetilde{\xi}$
plane an upper threshold exists. As discussed in the $n=2$ section,
this can be done by examining the limit of large $k$, in which the
annihilation process becomes indistinguishable from photon decay.  The
decay process is forbidden below the broken line given by
$\widetilde{\xi}=\widetilde{\eta}/2$ for $=\widetilde{\eta}>0$ and
$\widetilde{\xi}=\widetilde{\eta}$ for $\eta<0$.  Above this broken
line there is a lower threshold and no upper threshold.  Thus an upper
threshold exists below this broken line anywhere a lower threshold
exists.\footnote{
%-------------------------------------------------------
This analysis was carried out using the truncation discussed in
section\ref{kinematics}. At very high energies or at very large values
of the Lorentz violating parameters there will be deviations from this
behavior, but for practical purposes these deviations are not relevant
for our constraints.}
%-------------------------------------------------------

\begin{figure}[htb]
\vbox{ \vskip 8 pt \centerline{\includegraphics[width=5in]{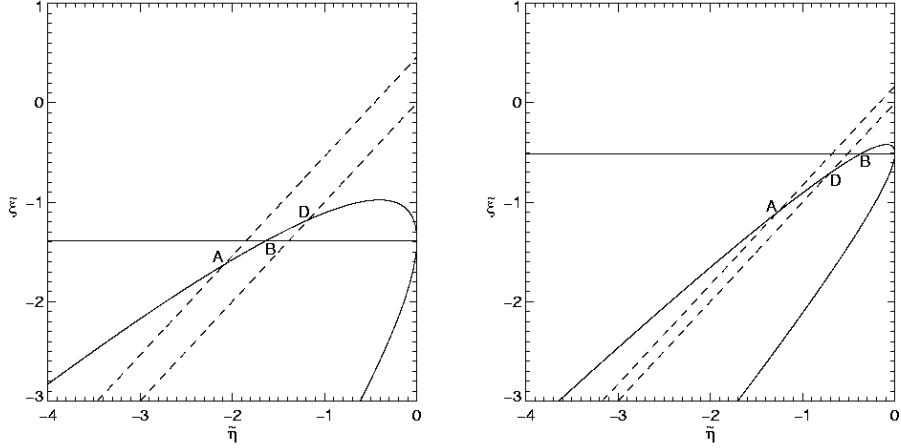}}
\caption{\label{fig:upperb2} The asymmetric curve corresponds to a
local maximum for all points on the segment from $A$ (where it crosses
the line $\widetilde{\xi}=\widetilde{\eta}+4/3\b^2$) to $B$ (where it
crosses $\widetilde{\xi}=-4/\b^2$ and becomes unphysical. It is the
global maximum only in the region below the diagonal, which it crosses
at $D$. If $\b<2$, $D$ lies above $B$ so the asymmetric configuration
is never a global maximum. If $\b>2$ then the asymmetric configuration
is the global maximum for every point on the segment between $D$ and
$B$.
\smallskip}
}
\end{figure}

The candidates for upper threshold contours are the sections of
$\b$-curves with $\b>1.5$ that satisfy the conditions for being a
local maximum. On the symmetric segment this imposes no restriction,
but on the asymmetric segment it requires that the curve lie below the
line $\widetilde{\xi}= \widetilde{\eta}+4/3\b^2$.  On the other hand,
we just argued that an upper threshold exists only below the diagonal
for negative $\eta$, which is a more restrictive condition. Moreover,
it can be checked that these sections of the $\b$-curves do not cross
anywhere in the region of upper thresholds, hence they are indeed the
upper threshold contours in that region.  Only for $\b>2$ does the
asymmetric section have a segment below the diagonal before leaving
the physical region $\widetilde{\xi}<-4/\b^2$, as illustrated in
Fig.~\ref{fig:upperb2}.
%-------------------------------------------------------------------
%-------------------------------------------------------------------
%
The regions of symmetric and asymmetric upper thresholds thus take the
form shown in Figure~\ref{fig:upperregions}.  The the boundary of the
lens shaped region next to the diagonal is determined by the curve
(\ref{eq:xijoin}) consisting of the points where the symmetric and
asymmetric segements join. The bottom of the lens meets the diagonal
at the $\b=2$ line.

%----------------------------------------------------------------

%----------------------------------------------------------------
%----------------------------------------------------------------
\end{document}